\documentclass[%
reprint,
superscriptaddress,
 amsmath,amssymb,
 aps,
 pra,
]{revtex4-2}

\usepackage{textgreek}
\usepackage{mathtools}
\usepackage{nicefrac}
\usepackage{siunitx}
\usepackage{multirow}
\usepackage[normalem]{ulem}
\usepackage{graphicx}% Include figure files
\usepackage{dcolumn}% Align table columns on decimal point
\usepackage{bm}% bold math
\usepackage{microtype}
\usepackage[
   colorlinks=true,
   citecolor=blue,
   linkcolor=blue,
   urlcolor=blue,
]{hyperref}
\usepackage{orcidlink}

\newcommand{\dmscr}{\delta \langle r_\mathrm{c}^2 \rangle}
\begin{document}

\title[Hyperfine Constants of 15 Energy Levels in \textsuperscript{169}Tm II]{Magnetic Dipole Hyperfine Constants of 15 Energy Levels in \textsuperscript{169}Tm II\\ Determined by Collinear Laser Fluorescence Spectroscopy}

\author{Hendrik Bodnar\orcidlink{0009-0005-3056-3124}}
\email{hbodnar@ikp.tu-darmstadt.de}
\affiliation{Institut f\"ur Kernphysik, Technische Universit\"at Darmstadt, 64289 Darmstadt, Germany}

\author{Kristian König\orcidlink{0000-0001-9415-3208}}
\affiliation{Institut f\"ur Kernphysik, Technische Universit\"at Darmstadt, 64289 Darmstadt, Germany}%
\affiliation{Helmholtz Forschungsakademie Hessen für FAIR, GSI Helmholtzzentrum für Schwerionenforschung, 64291 Darmstadt, Germany}
\author{Wilfried N\"ortersh\"auser\orcidlink{0000-0001-7432-3687}}
\affiliation{Institut f\"ur Kernphysik, Technische Universit\"at Darmstadt, 64289 Darmstadt, Germany}%
\affiliation{Helmholtz Forschungsakademie Hessen für FAIR, GSI Helmholtzzentrum für Schwerionenforschung, 64291 Darmstadt, Germany}

\begin{abstract}
We present measurements of the hyperfine structure, transition frequencies and relative intensities in 14 lines of $^{169}$Tm$^+$ ions starting from the $4f^{13}({}^2\!F^o_{7/2})6s_{1/2} (7/2, 1/2)_{3,4}^o$ levels. The magnetic dipole hyperfine ($A$) constant of all 15 involved ionic levels are determined, 14 of which have sub-MHz accuracy. For five levels no previous measurements were reported, while all others are in excellent agreement with the literature and improve on the previous accuracy by typically 1 - 2 orders of magnitude. We compare transition strengths and include improved transition frequencies. These results are useful for the quantum computing community, modelling of Tm$^+$ abundances in stellar spectra, as well as for an ongoing measurement program on short-lived Tm isotopes at ISOLDE/CERN.
\end{abstract}

\maketitle

\section{Introduction}
\label{sec:intro}
The atomic structure of the element thulium (Tm, $Z=69$), one of the rare-earth elements, is of interest for quantum computing, stellar observations, and also from a nuclear structure point of view. Thulium atoms combine multiple advantages of alkali atoms and alkaline-earth-like atoms due to their complex $4f$ electronic structure. They provide the possibility of robust qubit encoding in the $m_F = 0$ sublevels of the ground-state hyperfine doublet with large natural depolarization times, as in alkaline atoms. It also only has a single stable isotope, $^{169}$Tm. This is convenient, as there is no isotope-selection issue and the $I=1/2$ nuclear spin results in a clean and simple hyperfine structure. At the same time, they have suitable optical transitions for laser cooling \cite{Sukachev.2014,Provorchenko.2023} and the ability to store qubits in metastable states, like alkaline-earth elements. Thus, Tm was proposed for quantum simulations, and record coherence times of up to one minute with minimal external control were demonstrated, together with coherent transfer and storage of the qubit state in a metastable level \cite{Mishin.2025}. Neutral Tm is a well established system that is also used for atomic clocks \cite{Sukachev.2016,Golovizin.2019,Golovizin.2024} and Bose–Einstein condensates \cite{Khlebnikov.2019,Davletov.2020,Tsyganok.2023}.

The singly charged ion Tm$^+$ has similar advantages: the complex $4f$-shell structure leads to an abundance of accessible electronic states, including a number of long-lived states that can be exploited in the optical, metastable and ground-state qubit protocol ($\textit{omg}$). Here, too,  one ion species can be used for multiple tasks. Different qubit types utilize different states and are thus largely insensitive to the laser light required for another qubit type. This allows parallelizing laser cooling, memory storage and quantum-gate execution in distinct atom registers, using only a single ion species \cite{Allcock.2021}. Moreover, Tm$^+$ has laser-accessible transitions connecting a multitude of states with different total electronic angular momentum up to $J=9$ \cite{Biemont.1999}, which is favorable for higher order error detection in absorption-emission codes \cite{Shubham.2024, Aydin.2025}. While the feasibility of laser cooling open-$4f$-shell rare-earth ions had been discussed theoretically \cite{Lepers.2016}, its first experimental demonstration was achieved only recently in $^{169}$Tm$^+$ \cite{Mueller.2026b}.

The second line of interest in these systems is the extraction of nuclear properties of short-lived, neutron-deficient Tm isotopes. Such information is of particular interest around the $N=82$ nuclear shell closure at $^{151}$Tm and towards the proton drip line. The isotope $^{147}$Tm is a proton emitter, for which the charge radius behavior is not yet known. Laser spectroscopy can be used to measure the isotope shifts and the hyperfine structure splitting, from which $\dmscr$ and the magnetic dipole and nuclear quadrupole moments can be extracted, respectively \cite{Cheal.2010,Blaum.2013,Campbell.2016,Neugart.2017,Yang.2023}. The Tm isotope chain has been investigated in atomic transitions by in-source resonance ionization spectroscopy in the range $^{157-172}$Tm \cite{Alkhazov.1988} and $^{153-155}$Tm \cite{Barzakh.2000} at Gatchina. Recently, further results for $^{152m}$Tm, $^{153}$Tm, $^{154m}$Tm, and $^{169,170}$Tm have appeared from investigations at the SHIP experiment at the GSI Helmholtz Center in Darmstadt \cite{Weyrich.2026}, where neutral Tm atoms have been investigated in a gas cell using a three-step ionization scheme out of the atomic ground state. At CERN/ISOLDE, a program at the COLLAPS experiment has targeted the isotopic chain of Tm by collinear laser spectroscopy of Tm$^+$ ions \cite{Cheal_ISOLDE_LoI, Cheal_ISOLDE_Proposal}, which provides significantly higher accuracy for the determination of hyperfine structure splitting constants and the isotope shifts. 

Finally, Tm$^{+}$ lines are also of relevance to astrophysics because rare earth elements, though intrinsically among the least abundant species in stellar atmospheres, provide sensitive probes of heavy-element nucleosynthesis (e.g., neutron-capture processes) and appear as numerous, well-resolved spectral lines in the Sun \cite{Wang.2022} and in chemically peculiar, metal-poor and evolved stars, like Przybylski's Star (HD 101065) \cite{Cowley1998}. Reliable laboratory measurements of the HFS constants are essential for correctly modeling Tm II line shapes \cite{Kebapci.2024} in these spectra and support more accurate abundance analyses of stars.

Despite these interests, there is little information so far on the hyperfine structure splittings in the levels of $^{169}$Tm$^+$. 
Collinear fast-ion-beam laser spectroscopy was used in \cite{Mansour.1989} to determine the magnetic dipole hyperfine constant $A$ for eight fine-structure levels, some of them in tension with multiconfiguration Dirac-Hartree-Fock calculations \cite{Cheng.1985}. Only recently, additional measurements were reported in \cite{Kebapci.2024}, which corrected two of the $A$ parameters bringing them into better agreement with theory. In total, 40 Tm II lines in the wavelength range from 335--2345\,nm were measured using Fourier-transform spectrometry of emission lines from a hollow-cathode lamp, providing $A$ for 27 additional levels. Also, lately, the hyperfine $A$ constants for all levels relevant to establish the primary $313\,\mathrm{nm}$ and complementary $448/453\,\mathrm{nm}$ cycling transitions were reported, including additional lifetime measurements and Zeeman-resolved microwave hyperfine spectroscopy with kHz precision on one of the candidates for hosting a robust qubit \cite{Mueller.2026b}. The planned and ongoing experiments at ISOLDE/CERN motivated additional theoretical calculations of the magnetic dipole HFS constants $A$ and Landé $g$ factors for the levels of the ground $4 f^{13} 6 s$ and first excited $4 f^{13} 5d$ odd configurations in Tm II using a configuration interaction method with the random-phase approximation included \cite{Bondarev.2026}. 

Here, we report measurements on the rest-frame frequency and the hyperfine structure of the stable isotope $^{169}$Tm in several UV and near-UV transitions of the singly charged ion that are potentially of interest. The relative fluorescence signal intensity after collinear laser excitation was also determined since this determines the sensitivity for the online measurements at ISOLDE/CERN. Preliminary values of our hyperfine structure results have already been used in \cite{Mueller.2026b} to extract the magnetic dipole constants of other levels, demonstrating the utility of our results also for the quantum computing community.

\section{Experimental Setup}
\label{sec:Experiment}

Table\,\ref{tab:Transitions} lists all investigated transitions. These connect 15 different energy levels and have been chosen as they either had transition probabilities of reasonable strength assigned to them in the NIST database \cite{Kramida.1999} or, those marked with $^{*}$ in the table, had high transition probabilities in the Kurucz database \cite{Kurucz.1995}. All of them start from either the ground state or the first excited state at $236.95$\,cm$^{-1}$.

\begin{table}[tb]
    \centering
    \caption{The transitions in singly ionized thulium studied in this work. The wavelengths marked with $^{*}$ were taken from the Kurucz database \cite{Kurucz.1995}. All other information originates from the NIST database \cite{Kramida.1999}. The labels l and u denote lower and upper, respectively. The configuration of the two low-lying odd-parity energy levels is $4f^{13}({}^2\!F^o_{7/2})6s_{1/2} (7/2, 1/2)^o$ with $J=3$ and $J=4$ for the ground state and the $236.95$\,cm$^{-1}$ energy level, respectively.}
\begin{ruledtabular}
\begin{tabular}{S[table-format=3.2]@{}l S[table-format=3.2] S[table-format=5.2] c c@{\hspace{2em}} l}
\multicolumn{2}{c}{Wavelength} & {$E_{\mathrm{l}}$}  &  {$E_{\mathrm{u}}$} & $J_{\mathrm{l}}$  &  $J_{\mathrm{u}}$ & Upper \\
\multicolumn{2}{c}{(nm)} & {(cm$^{-1}$)}
                         & {(cm$^{-1}$)}
                         &  &  & {Configuration} \\
\hline
248.01 & $^{*}$ & 236.95 & 40545.25 & 3 & 3 & $4f^{12}\,5d^{2}$ \rule{0mm}{3.5mm}\\
250.91 & $^{*}$ &      0 & 39843.24 & 4 & 4 & $4f^{12}\,5d6s$ \\
313.13 &        &      0 & 31926.82 & 4 & 5 & $4f^{12}\,5d6s$ \\
336.26 &        & 236.95 & 29967.17 & 3 & 2 & $4f^{12}\,5d6s$ \\
342.51 &        & 236.95 & 29424.98 & 3 & 3 & $4f^{12}\,5d6s$ \\
344.15 &        & 236.95 & 29285.72 & 3 & 2 & $4f^{13}\,6p$ \\
345.37 &        & 236.95 & 29183.39 & 3 & 4 & $4f^{12}\,5d6s$ \\
346.22 &        &      0 & 28875.12 & 4 & 5 & $4f^{12}\,5d6s$ \\
370.03 &        & 236.95 & 27254.42 & 3 & 4 & $4f^{12}\,5d6s$ \\
370.14 &        &      0 & 27009.39 & 4 & 4 & $4f^{12}\,5d6s$ \\
376.13 &        &      0 & 26578.77 & 4 & 3 & $4f^{12}\,6s^{2}$ \\
376.19 &        &      0 & 26574.66 & 4 & 4 & $4f^{12}\,5d6s$ \\
379.58 &        & 236.95 & 26574.66 & 3 & 4 & $4f^{12}\,5d6s$ \\
384.80 &        &      0 & 25980.02 & 4 & 3 & $4f^{13}\,6p$ \\
\end{tabular}
\label{tab:Transitions}
\end{ruledtabular}
\end{table}

The measurements were performed at the COllinear Apparatus for Laser spectroscopy and Applied science (COALA) at TU Darmstadt.
The entire setup with its core features is presented and explained in detail in Ref.\,\cite{Koenig.2020}. Only a brief summary is provided here.

Thulium ions are created in a surface ionization source \cite{Raeder.2011,Frommgen.2013}. A graphite crucible is filled with small solid thulium pieces and heated by a current of up to 100\,A. The heating currents are $35\,\mathrm{A}$ to $50\,\mathrm{A}$ ($\approx$ $1330\,^\circ\mathrm{C}$ - $1770\,^\circ\mathrm{C}$) \cite{Frommgen.2013} to generate ion beams of $200\, \mathrm{pA}$ up to $2.5\, \mathrm{nA}$ at the beam dump at the end of the collinear beamline. 
The ion source is floated at 20\,kV, accelerating the ions towards the rest of the beamline on ground potential. The ion beam is guided to the interaction region using electrostatic ion optics. Three beam diagnostic stations are installed along the beamline. Each allows to measure the ion current arriving at a Faraday cup and to image the shape and position of the ion and laser beams on a phosphorus screen.

The laser-ion interaction takes place in the fluorescence detection region (FDR) located between the last two diagnostic stations. It can be floated on an individual potential and is composed of two successive chambers. The first one contains a mirror system while in the second, a lens-based detection is installed. The mirror system has an elliptical cross section in the plane perpendicular to the beam direction with the ion beam in the first focal point and a length of $8\,\mathrm{cm}$ in beam direction. On top of it, two photomultiplier tubes (PMTs) are installed along the second focal line of the mirror system \cite{Maass.2020,Koenig.2020}. A third PMT is mounted at the focal point of the lens system \cite{Mueller.2024}. This channel detects less laser-generated background, but it also has a lower overall efficiency. For short measurement times and relatively high ion currents, it provides a higher signal-to-noise ratio (SNR) \cite{Mueller.2024}.

A Kepco BOP 500M is used for applying a scanning voltage ($U_{\text{FDR}}$) to the FDR for Doppler-tuning, providing a full range of $1\,\mathrm{kV}$. Considering the relativistic Doppler factor of \textsuperscript{169}Tm\textsuperscript{+}, this translates to more than 10\,GHz in frequency space, depending on the transition, being more than sufficient to record the peaks of the full HFS, spanning up to $8\,\mathrm{GHz}$ at maximum. 

For laser excitation of the transitions with wavelengths in the $248 - 385\,\mathrm{nm}$ range, Ti:Sapphire (Ti:Sa, Matisse 2 TS, Sirah Lasertechnik) and dye lasers (Matisse 2 DS), both pumped by Millennia eV Nd:YVO$_4$ lasers (Spectra Physics), are used in various combinations with second-harmonic generation cavities (WaveTrain 2, Spectra Physics).
To access the transitions from $336 - 385\,\mathrm{nm}$, the Ti:Sa is combined with a Wavetrain, while the shortest wavelengths, around $250\,\mathrm{nm}$, is produced by frequency quadrupling the Ti:Sa light with two Wavetrains in sequence. Only the $313.13\,\mathrm{nm}$ transition requires the dye laser operated with Rhodamine B and paired with a Wavetrain. 
Frequency stabilization is realized with a Wavemeter (HighFinesse WSU10) \cite{Verlinde.2020,Koenig.2020b} and typically limits drifts to less than $0.4\,$MHz/h.

\section{Experimental Results and Discussion}
\label{sec:results}

\subsection{Measurements and Analysis}
\label{ssec:measurements}
All measurements were performed at an ion beam energy of about 20\,keV. The high voltage at the ion source is actively stabilized to a precision of 0.06\,V, by applying a secondary, adjustable high-voltage correction \cite{Koenig.2024}. For transitions in the wavelength range $336 - 385\,$nm, the laser beam was transported to the COALA beamline with a fiber, whereas for the deep UV transitions air transmission of the beam was applied. All hyperfine structure resonances were recorded using Doppler tuning with a typical step width of about 1\,V, corresponding to roughly 3--4\,MHz. The laser beam intensity was recorded before each scan and the ion beam intensity, laser frequency and acceleration voltage were recorded continuously during every measurement. 

For each spectrum, several scans across the main resonances (typically about 20) were added, while, in many cases, less intense hyperfine components were scanned more often, until a similar SNR was obtained. Therefore, the signal rate was always normalized to the number of scans. In these spectra, the individual peaks were scanned in consecutive blocks. Both chambers of the FDR have been used and evaluated separately. Results from both chambers agreed within statistical uncertainties and were averaged.

The relative intensity of the different lines is an important information for on-line measurements of rare species, since a weak transition requires higher production yields to obtain a given SNR within a fixed time span than a transition with a higher intensity. To obtain such information, the $384.8\,$nm line was chosen as a reference for all other transition and measured either before or after the measurements of another transition to minimize the impact of differing experimental conditions on the respective day. The normalization procedure is explained below.

One consequence of this is an imbalance in the number of measurements for each energy level. The ground state appeared in $108$, the first excited state in $16$, the $25980.02\,\mathrm{cm}^{-1}$ state in $94$ and the rest in around $1 - 5$ measurements. This directly impacts the statistical uncertainties of the results. 

\begin{figure*}
    \centering
    \includegraphics[width=1\linewidth]{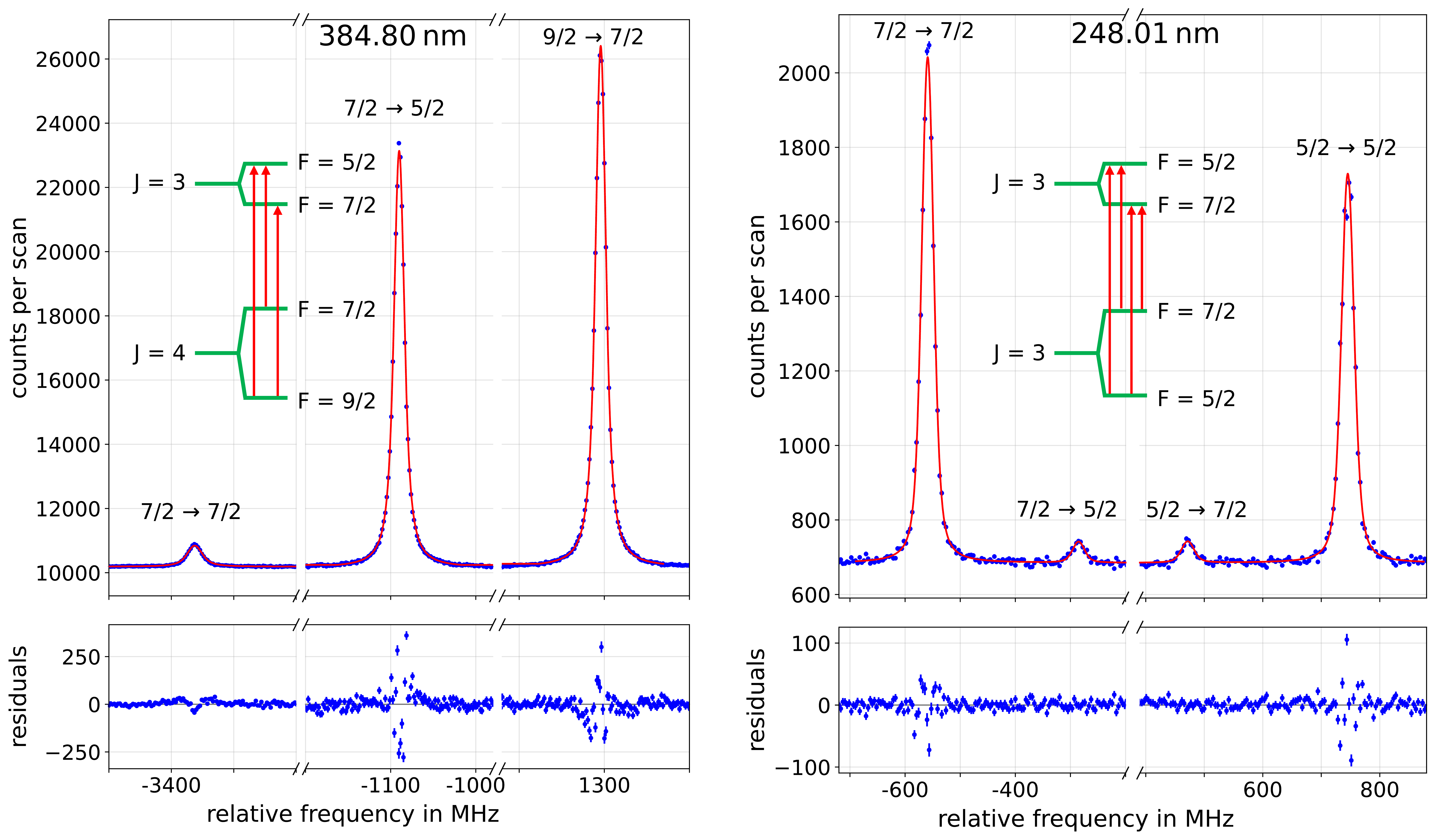}
    \caption{Example spectra of the $4f^{13}6s_{\nicefrac{1}{2}} (\nicefrac{7}{2}, \nicefrac{1}{2})^o_4 \rightarrow 4f^{13}6p$ $J = 3$ transition at $384.8$\,nm on the left and the $4f^{13}6s_{\nicefrac{1}{2}} (\nicefrac{7}{2}, \nicefrac{1}{2})^o_3 \rightarrow 4f^{12}\,5d^{2}$ $J = 3$ transition at $248.01\,$nm on the right. The $x$-axes show the frequency relative to the center of gravity of the respective transition in MHz. The $y$-axes display the counts per scan. In the $384.8\,$nm measurement, the smallest peak was scanned more often to obtain a similar SNR as the other peaks. Thus, the $y$-axis has been normalized to the number of scans. The blue points are the data points including error bars and the red line represents the fit. Residuals between fit and data points are indicated in the lower frames. The insets show the hyperfine level scheme of the respective transition (not to scale). Above each peak, the individual hyperfine transition is indicated by the corresponding change of total angular momentum $F \rightarrow F^{\prime}$.}
    \label{fig:Example_Spectrum}
\end{figure*}

The hyperfine structure in the $384.8\,\mathrm{nm}$ line ($J = 4 \rightarrow J = 3$) (reference) and the $248.01\,$nm line ($J = 3 \rightarrow J = 3$) are shown in Fig.\,\ref{fig:Example_Spectrum} as typical example of the recorded spectra. The $x$-axis represents the Doppler-tuned frequency relative to the center of gravity of the transition, while the $y$-axis displays the observed PMT events, normalized to the number of scans. Uncertainties of the individual data points are based on counting statistics. The inset shows the level scheme of the transitions, including the hyperfine structure. The data in both measurements was taken in the first FDR chamber.

At the begin of the measurement campaign, the reference line was also measured in collinear and anticollinear geometry quasi-simultaneously to obtain the transition frequency in the rest-frame of the ion. The result is reported in Sec.\,\ref{ssec:transfreqs} and has been used to obtain the ion beam energy on each measurement day from the anticollinear measurements of the reference transition, being independent from the high-voltage determination and a direct measure of the beam energy.

For fitting, a Voigt lineshape was used for each hyperfine component, taking into account the Lorentzian component from the natural linewidth as well as a residual Doppler contribution. The line-position with respect to the center of gravity (cg) was calculated from $\nu = \nu_{\mathrm{cg}} + \nicefrac{1}{2} (A_{\mathrm{u}}\, C(F_{\mathrm{u}}) - A_{\mathrm{l}}\, C(F_{\mathrm{l}}))$ with $A_{\mathrm{l}}$, $F_{\mathrm{l}}$  and $A_{\mathrm{u}}$, $F_{\mathrm{u}}$ being the hyperfine constants and total angular momentum of the lower and the upper level of the transition, respectively. $C\left(F\right)=F(F+1)-I(I+1)-J(J+1)$ is the Casimir factor of the respective state. The free parameters of the fit are $\nu_{\mathrm{cg}}$, $A_{\mathrm{l}}$, $A_{\mathrm{u}}$, the individual peak heights, the common Lorentzian-, and Gaussian linewidth parameters and the offset. The fit result is represented as a red line and the residuals are plotted in the lower frame. Several other fit functions (Lorentzian, Gaussian, asymmetric Voigt) were also used for fitting with a small variation of the reduced $\chi^2$ but with marginal influence on the obtained $A$-constants ($<5\,\mathrm{kHz}$), being only a fraction of its statistical uncertainty. A small structure in the residuals remains, almost independent of the choice of fit function. Thus, the final values were taken from fits with the Voigt lineshape.

\subsection{Hyperfine Constants}
\label{ssec:HyperfineConstants}
The fit of a single measurement yielded $A$-values with their associated statistical uncertainty. The influence of fluctuating laser frequencies and acceleration voltages to this scan was then taken into account based on the simultaneously measured frequencies and high voltages during the scan: their standard variations were calculated, the voltage fluctuations were transformed into frequency space and both contributions were added quadratically. For a conservative error estimation, it was assumed that the peaks on the two sides of $\nu_{\mathrm{cg}}$ are affected in opposite directions. This presented the statistical uncertainty of one measurement. In cases with multiple measurements per energy level, a weighted average was performed. The final statistical uncertainty was taken as the larger of either the standard error of the mean or the standard error of the weighted average.

The $A$ factors of the two highest energy levels have larger statistical uncertainties due to issues with the laser stabilization at these wavelengths. A $10\,\mathrm{MHz}$ standard deviation of the laser frequency for the $248.01\,\mathrm{nm}$ measurements was assumed. It was estimated from the quality of the spectrum and the width of the peaks directly. For the $250.91\,\mathrm{nm}$ measurements, the peaks had even visibly shifted over the course of the measurements leading to a lineshape poorly described by a Voigt or one of the other used lineshape functions and with significantly broader width, which actually coincided with the recorded frequency fluctuations of about $\pm 30$\,MHz during the scan. Thus, the statistical uncertainty was in this single case conservatively estimated as $60\,\mathrm{MHz}$.

Looking at potential systematic uncertainties, most factors cancel out or are significantly dampened by the fact that the $A$-constant is obtained purely from frequency differences of the peaks within a single scan. Thus, systematic uncertainties arising from the uncertainty of the laser frequency are of the order of a few kHz and can be completely neglected. The uncertainty of the beam energy was estimated to be at most $1.4\,$eV, see \ref{ssec:transfreqs} for the procedure. This translates to a $0.004\,\%$ uncertainty of the $A$-constants The second most important factor comes from the scanning voltage applied to the FDR. Due to the nature of the construction of both photon collection systems (elliptical mirror and lens system), there exist field penetrations in the FDR which affect the ion beam energy while the resonance is measured. The magnitude of the field penetration was determined \cite{Koenig.2024} but has associated uncertainties leading to a $0.003\,\%$ systematic uncertainty in all $A$-constants. The systematic uncertainties are represented in Table\,\ref{tab:HFS Table} as the second uncertainty.

Listed in the last column of the table are literature values \cite{Kebapci.2024,Mansour.1989}. The agreement is well within $1\,\sigma$ uncertainty, but our vales are improved by 1--2 orders of magnitude in accuracy. For the ground state and the first excited state, a comparison with recent theoretical calculations \cite{Bondarev.2026} is possible. The configuration interaction method with the random-phase approximation yields results reasonably close to experiment. In the ground state the agreement is excellent with a difference of only $0.96\,$MHz, whereas it amounts to about 42\,MHz for the $236.95\,$cm$^{-1}$ state. The larger discrepancy is explained by the opposite orientation of the $s$-electron spin and the total angular momentum of the electron hole in the $4f^{13}\,(^2\!F_{\nicefrac{7}{2}})$ configuration, which leads to partial cancellation of their contributions to $A$ and makes the $A$ factors more difficult to calculate.

\begin{table}[tb]
    \centering
    \caption{The magnetic-dipole hyperfine constant $A$ for the energy levels $E$ under investigation from Table\,\ref{tab:Transitions}. The first uncertainty of $A$ is the statistical uncertainty, the second one represents the systematic uncertainty originating from field penetrations into the florescence detection region (FDR) as discussed in the text. The $J$=4 ground state appeared in $108$, the first excited state in $16$, the $25980.02\,\mathrm{cm}^{-1}$ state in $94$ and the others in around $1 - 5$ measurements. The literature values are taken from \cite{Kebapci.2024}, except the value marked with $^{*}$ reported in \cite{Mansour.1989}.}
    \setlength{\tabcolsep}{1pt}
    \begin{ruledtabular}
    \begin{tabular}{S[table-format=5.2] l S[table-format=-4.3{\, (000)\, (0)}, table-align-text-after=false] S[table-format=-4.1{\,(0.0)}, table-align-text-after=false]}
        {\rule{0mm}{3.5mm}$E$ (cm$^{-1}$)} & {Configuration} & {$A$ (MHz)} & {Lit. (MHz)} \\
    \hline
            \rule{0mm}{3.5mm}0    & {($4f^{13}6s)_{J=4}$}   & -1039.04 {\,(2)\,(6)}   & -1038.3\,(1.6)      \\
          236.95 & {($4f^{13}6s)_{J=3}$}   & 294.73   {\,(6)\,(2)}    & 293\,(6)            \\
        25980.02 & {($4f^{13}6p)_{J=3}$}   & -654.40  {\,(2)\,(4)}   & -654\,(4)           \\
        26574.66 & {($4f^{12}5d6s)_{J=4}$} & -206.64  {\,(4)\,(1)}    & -206.4\,(1.9)       \\
        26578.77 & {($4f^{12}6s^2)_{J=3}$} & -136.79  {\,(4)\,(1)}    & -136.6\,(1.0)$^{*}$ \\
        27009.39 & {($4f^{12}5d6s)_{J=4}$} & -820.70  {\,(16)\,(5)}  & -816\,(5)           \\
        27254.42 & {($4f^{12}5d6s)_{J=4}$} & -502.48  {\,(25)\,(3)}  & -503\,(6)           \\
        28875.12 & {($4f^{12}5d6s)_{J=5}$} & -661.13  {\,(9)\,(4)}   &                     \\
        29183.39 & {($4f^{12}5d6s)_{J=4}$} & -482.67  {\,(10)\,(3)}  & -481\,(5)           \\
        29285.72 & {($4f^{13}6p)_{J=2}$}   & -279.11  {\,(15)\,(2)}   &                     \\
        29424.98 & {($4f^{12}5d6s)_{J=3}$} & -175.78  {\,(16)\,(1)}   & -175\,(6)           \\
        29967.17 & {($4f^{12}5d6s)_{J=2}$} & -1044.87 {\,(21)\,(6)}  & -1045\,(8)          \\
        31926.82 & {($4f^{12}5d6s)_{J=5}$} & -718.71  {\,(22)\,(4)}  &                     \\
        39843.24 & {($4f^{12}5d6s)_{J=4}$} & -699.14  {\,(440)\,(4)} &                     \\
        40545.25 & {($4f^{12}5d^2)_{J=3}$} & -78.26   {\,(53)\,(1)}   &                     \\
    \end{tabular}
    \end{ruledtabular}
    \label{tab:HFS Table}
\end{table}

\subsection{Transition Frequencies}
\label{ssec:transfreqs}
Transition frequencies were determined for all lines with higher accuracy than available in literature. The reference transition was used for a determination of the ion beam energy as described in \cite{Konig.2021}. Though the applied acceleration voltage is recorded and stabilized, we do not know the precise conditions in the ion source and the exact starting potential of the ions. Therefore, the $F=\nicefrac{7}{2} \rightarrow F=\nicefrac{5}{2}$ hyperfine component of the $384.8\,\mathrm{nm}$ line was used as a beam-energy reference. For this resonance, the absolute transition frequency was determined via a collinear/anticollinear measurement, which minimizes the starting energy dependency, analog to our previous measurements, \textit{e.g.}, in Ba$^+$, Ca$^+$, Sr$^+$ and C$^{4+}$ \cite{Imgram.2019,Muller.2020,Palmes.2026,Imgram.2023,Koenig.2026}. In this case, the laser was stabilized to a Menlo Systems frequency comb (FC1500-250-WG) instead of the wavemeter. The two laboratory frequencies of the collinear $\nu_c$ and the anticollinear $\nu_a$ laser, with the respective resonances appearing at identical post-acceleration voltages, provide the transition frequency $\nu_0$ in the ions rest frame. The determined frequency of the hyperfine component and that of the center of gravity of the transition are shown in Table \ref{tab:transfreqs}.

Five pairs of collinear/anticollinear measurements were performed leading to a statistical uncertainty of $62\,$kHz. This is defined as the standard error of the mean after a weighted average of all pairs. The systematic uncertainty is completely dominated by the laser frequency contribution. At that time, the frequency comb was suffering from an aging pumping diode, leading to inconsistent readings of the beat frequency. In Ref.\,\cite{Palmes.2026}, this effect was characterized in detail and an uncertainty of $600\,$kHz was reported. Because only five measurement pairs were available in the Tm case, this work could not retrospectively characterize the effect independently with comparable confidence. We therefore conservatively assign an uncertainty of $1.2\,$MHz, corresponding to twice the value reported in Ref.\,\cite{Palmes.2026}. Factors such as the photon recoil and slightly differing ion beam energies between collinear and anticollinear measurements were considered in the calculations. The remaining contributions from the choice of lineshape ($< 10\,$kHz) and the laser-laser and laser-ion overlap ($32\,$kHz) can be neglected.

Experimental data on transition frequencies of the other transitions were, to our knowledge, previously limited to identification of their wavelengths in copper arc spectra \cite{Meggers.1975, Reader.1980}. Both reports do not provide uncertainties for the relevant wavelengths. Reader \textit{et al}.~\cite{Reader.1980} state that wavelengths tabulated to three decimal places have uncertainties below $0.001\,\text{\AA}$. All transitions considered here are, however, tabulated to only two decimal places. Over the wavelength range investigated in this work, $0.001\,\text{\AA}$ corresponds to approximately $200 - 500\,$MHz. We have determined all frequencies based on the wavemeter readings with wavemeter uncertainty of 40\,MHz for the frequency quadrupled light and 20\,MHz for the frequency doubled light. 

The transition frequencies are compiled in Table \ref{tab:transfreqs}. The wavelengths in the first column serve only as line identifiers, whereas the second and third columns give the measured rest-frame frequencies and the corresponding wavenumbers, respectively. The uncertainties in the table are the combined statistic and systematic uncertainties, which is dominated by the wavemeter's systematic uncertainty. The beam energy measured with the reference transition was used to estimate the starting potential uncertainty of the ions, resulting in a value of at most $1.4\,\mathrm{V}$ (70\,ppm). This, added quadratically to the wavemeter uncertainty, adds only a small additional contribution of up to $6\,$MHz on top. Other contributions, like the field penetration inside the FDR are $<1$\,MHz and, thus, completely negligible compared to the other systematics. 

\begin{table}[tb]
    \centering
    \caption{The rest-frame frequencies of all transitions measured in this work. The transition marked with the label hf is the $F = \nicefrac{7}{2} \rightarrow F = \nicefrac{5}{2}$ hyperfine component of the $384.8\,$nm transition. Here, the uncertainties are given separately, first statistical, then systematic. For all other transitions, the uncertainties are combined. The frequencies are also converted to wavenumbers.}
    \begin{ruledtabular}
    \begin{tabular}{S S[table-format=10.2{\, (00)\, (0.00)}, table-align-text-after=false] S[table-format=5.6{\, (00)\, (00)}, table-align-text-after=false]}
        {$\lambda$} & {Frequency} & {Wavenumber} \\
        {in nm} & {in MHz} & {in cm$^{-1}$} \\
    \hline
        {$384.80_{\mathrm{hf}}$} & 778860434.97\,(6)\,(1.20) & 25979.987628\,(2)\,(40) \\
        384.80 & 778861723.7\,(1.3) & 25980.030616\,(42) \\
        379.58 & 789584773\,(24) & 26337.7130\,(14) \\
        376.19 & 796688316\,(24) & 26574.6617\,(14) \\
        376.13 & 796811670\,(25) & 26578.7764\,(14) \\
        370.14 & 809721431\,(25) & 27009.3996\,(14) \\
        370.03 & 809963558\,(25) & 27017.4762\,(14) \\
        346.22 & 865654181\,(25) & 28875.1154\,(14) \\
        345.37 & 867792571\,(25) & 28946.4444\,(14) \\
        344.15 & 870860253\,(25) & 29048.7712\,(14) \\
        342.51 & 875035061\,(25) & 29188.0278\,(14) \\
        336.26 & 891289685\,(26) & 29730.2237\,(14) \\
        313.13 & 957142145\,(26) & 31926.8254\,(15) \\
        250.91 & 1194470760\,(45) & 39843.2558\,(15) \\
        248.01 & 1208412584\,(42) & 40308.3050\,(14) \\
    \end{tabular}
    \end{ruledtabular}
    \label{tab:transfreqs}
\end{table}

\subsection{Relative Line Intensities}
\label{ssec:intensities}
To estimate the sensitivity of the different transitions for on-line measurements of rare species, their signal strengths, defined as the sum of the observed heights of all hyperfine structure peaks, were compared. A spectrum of the reference transition was taken just before or after the investigation of each line. Most day-to-day variations in the experimental conditions were compensated by this approach. Smaller variations of the ion beam current $I$ and laser power $P$  are additionally corrected linearly by applying the following normalization procedure to the signal strength $S$

\begin{equation}
    S_{x, \mathrm{norm}} = \frac{S_x}{S_{\mathrm{ref}}} \frac{I_{\mathrm{ref}}}{I_x} \frac{P_{\mathrm{ref}} \, \nu_x}{\nu_{\mathrm{ref}} \, P_x } \frac{ \epsilon_{\mathrm{ref}}}{\epsilon_x}.
\end{equation}

The subscripts ref and $x$ refer to the measurement of the reference transition and the transition under investigation, respectively. We note that the laser power is transformed into the laser photon flux ($N=P/h\nu$). The correction includes also the wavelength-dependent quantum efficiency of the PMT detectors $\epsilon$. Since only small differences in the mentioned quantities are present in the corresponding $x$-measurement/reference measurement pairs and the laser power was far below saturation, a linear scaling is justified. 

\begin{figure}
    \centering
    \includegraphics[width=1\linewidth]{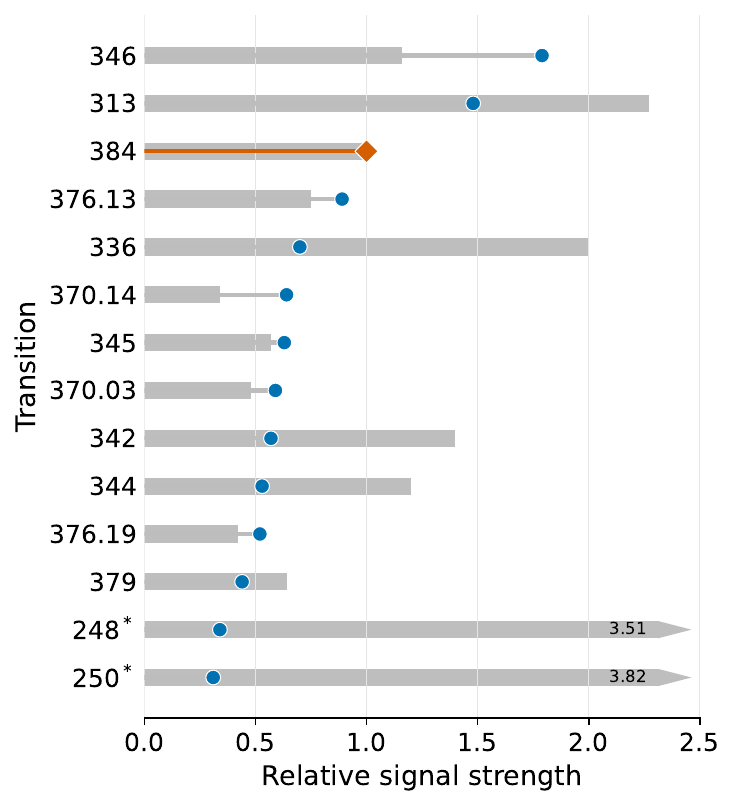}
    \caption{The signal strength of all measured transitions normalized to the $384.8\,\mathrm{nm}$ transition. The gray bars correspond to the ratio of Einstein $A_{21}$ values as listed in \cite{Kramida.1999, Kurucz.1995}. For the $248.01\,$nm and the $250.91\,$nm transitions, they are partially out of frame, extending to 3.51 and 3.82, respectively.}
    \label{fig:intensities}
\end{figure}

The overlap between the ion and laser beams is also of high importance when evaluating the intensity in the recorded spectrum. The laser beams originating from a fiber had the same trajectory between $x$-measurement and reference measurement. For the two deep UV lines the laser beams were adjusted to match the reference laser beam, so discrepancies are assumed to be negligible. The results are depicted in Fig.\,\ref{fig:intensities} and Table\,\ref{tab:ints}. 

The normalized fluorescence signals provide a semi-quantitative ranking of the transitions under our experimental conditions. The strongest signals were observed for the $346.22\,\mathrm{nm}$, the $313.13\,\mathrm{nm}$ and the $384.80\,\mathrm{nm}$ lines. Those are well suited for on-line measurements, where high sensitivity is an important requirement for isotopes with low production rates. All of these transitions also benefit from $\Delta J = 1$, as those transitions display fewer hyperfine transitions than a comparable $\Delta J = 0$ transition, making identification and fitting easier in the evaluation afterwards.

Also shown are the ratio of Einstein $A_{21}$ values \cite{Kramida.1999,Kurucz.1995} compared to the reference value. For wavelengths above $350\,$nm, the $A_{21}$ values are of the same order of magnitude as the relative intensities. The $346.22\,\mathrm{nm}$ line is a positive stand-out with its relative signal $\sim 1.5$ times higher than expected from the $A_{21}$ values. On the other hand, the deep UV, the $336.26\,\mathrm{nm}$, $342.51\,$nm and $344.15\,\mathrm{nm}$ transitions were weaker than expected.

\begin{table}[tb]
    \centering
    \caption{The normalized signal strength of each transition alongside the ratio of Einstein $A_{21}$ values \cite{Kramida.1999} relative to the reference. For the $248.01\,$nm and the $250.91\,$nm transitions the $A_{21}$ values are taken from \cite{Kurucz.1995}.}
    \setlength{\tabcolsep}{10pt}
    \begin{ruledtabular}
    \begin{tabular}{S S S}
        {Wavelength} & {Relative} & {Ratio of} \\
        {in nm} & {Signal Strength} & {$A_{21}$ values} \\
    \hline
        248.01 & 0.34 & 3.51 \\
        250.91 & 0.31 & 3.82 \\
        313.13 & 1.48 & 2.27 \\
        336.26 & 0.70 & 2.00 \\
        342.51 & 0.57 & 1.40 \\
        344.15 & 0.53 & 1.20 \\
        345.37 & 0.63 & 0.57 \\
        346.22 & 1.79 & 1.16 \\
        370.03 & 0.59 & 0.48 \\
        370.14 & 0.64 & 0.34 \\
        376.13 & 0.89 & 0.75 \\
        376.19 & 0.52 & 0.42 \\
        379.58 & 0.44 & 0.64 \\
        384.80 & 1    & 1    \\
    \end{tabular}
    \end{ruledtabular}
    \label{tab:ints}
\end{table}

\section{Conclusion}
\label{sec:conclusion}
We measured 14 lines in $^{169}$Tm$^+$ and extracted the magnetic dipole hyperfine constants for all 15 levels involved, with total uncertainties of typically less than $1\,$MHz. Five of these have no prior literature values assigned to them, whereas for all others, our results agree within the combined $1\,\sigma$ uncertainties and improve the precision by 1 -- 2 orders of magnitude. These values will be supportive for modeling Tm II lines in solar spectra as well as for the development of quantum computing schemes in these ions. 

The rest frame frequencies were determined for all transitions to an accuracy of at least an order of magnitude better than previously known, with the $384.8\,$nm line being the most precise at $1.3\,$MHz uncertainty.

A comparison of fluorescence signals normalized to nearby measurements of the $384.80\,$nm reference transition identifies the $346.22\,\mathrm{nm}$, the $313.13\,\mathrm{nm}$ and the $384.80\,\mathrm{nm}$ lines as promising candidates for collinear laser spectroscopy of low-yield neutron-deficient Tm isotopes at comparable experimental conditions.
The results provide spectroscopic input for the ongoing COLLAPS measurement program and for investigations of laser cooling and quantum-control schemes in $^{169}$Tm$^+$ as demonstrated in \cite{Mueller.2026b}.

\section{Data Availability}
The data that support the findings of this article, will be made openly available~\cite{data}.  

\begin{acknowledgments}
We are grateful for inspiring conversations with B.~Cheal, L.~Vázquez Rodríguez and T.~Lellinger that provided an initiating impetus for this research and with P.~Müller regarding the quantum computing applications of Tm II. We acknowledge support by the Deutsche Forschungsgemeinschaft (DFG, German Research Foundation) under Project Number NO789/4-1 and -- Project-ID 279384907 -- SFB 1245, and by the Bundesministerium für Forschung, Technologie und Raumfahrt (BMFTR, Federal Ministry of Research, Technology and Space) under Contract Nos. 05P21RDFN1 and 05P24RD4.
\end{acknowledgments}


\begin{thebibliography}{50}%
\makeatletter
\providecommand \@ifxundefined [1]{%
 \@ifx{#1\undefined}
}%
\providecommand \@ifnum [1]{%
 \ifnum #1\expandafter \@firstoftwo
 \else \expandafter \@secondoftwo
 \fi
}%
\providecommand \@ifx [1]{%
 \ifx #1\expandafter \@firstoftwo
 \else \expandafter \@secondoftwo
 \fi
}%
\providecommand \natexlab [1]{#1}%
\providecommand \enquote  [1]{``#1''}%
\providecommand \bibnamefont  [1]{#1}%
\providecommand \bibfnamefont [1]{#1}%
\providecommand \citenamefont [1]{#1}%
\providecommand \href@noop [0]{\@secondoftwo}%
\providecommand \href [0]{\begingroup \@sanitize@url \@href}%
\providecommand \@href[1]{\@@startlink{#1}\@@href}%
\providecommand \@@href[1]{\endgroup#1\@@endlink}%
\providecommand \@sanitize@url [0]{\catcode `\\12\catcode `\$12\catcode `\&12\catcode `\#12\catcode `\^12\catcode `\_12\catcode `\%12\relax}%
\providecommand \@@startlink[1]{}%
\providecommand \@@endlink[0]{}%
\providecommand \url  [0]{\begingroup\@sanitize@url \@url }%
\providecommand \@url [1]{\endgroup\@href {#1}{\urlprefix }}%
\providecommand \urlprefix  [0]{URL }%
\providecommand \Eprint [0]{\href }%
\providecommand \doibase [0]{https://doi.org/}%
\providecommand \selectlanguage [0]{\@gobble}%
\providecommand \bibinfo  [0]{\@secondoftwo}%
\providecommand \bibfield  [0]{\@secondoftwo}%
\providecommand \translation [1]{[#1]}%
\providecommand \BibitemOpen [0]{}%
\providecommand \bibitemStop [0]{}%
\providecommand \bibitemNoStop [0]{.\EOS\space}%
\providecommand \EOS [0]{\spacefactor3000\relax}%
\providecommand \BibitemShut  [1]{\csname bibitem#1\endcsname}%
\let\auto@bib@innerbib\@empty
%</preamble>
\bibitem [{\citenamefont {Sukachev}\ \emph {et~al.}(2014)\citenamefont {Sukachev}, \citenamefont {Kalganova}, \citenamefont {Sokolov}, \citenamefont {Fedorov}, \citenamefont {Vishnyakova}, \citenamefont {Akimov}, \citenamefont {Kolachevsky},\ and\ \citenamefont {Sorokin}}]{Sukachev.2014}%
  \BibitemOpen
  \bibfield  {author} {\bibinfo {author} {\bibfnamefont {D.}~\bibnamefont {Sukachev}}, \bibinfo {author} {\bibfnamefont {E.}~\bibnamefont {Kalganova}}, \bibinfo {author} {\bibfnamefont {A.}~\bibnamefont {Sokolov}}, \bibinfo {author} {\bibfnamefont {S.}~\bibnamefont {Fedorov}}, \bibinfo {author} {\bibfnamefont {G.}~\bibnamefont {Vishnyakova}}, \bibinfo {author} {\bibfnamefont {A.}~\bibnamefont {Akimov}}, \bibinfo {author} {\bibfnamefont {N.}~\bibnamefont {Kolachevsky}},\ and\ \bibinfo {author} {\bibfnamefont {V.}~\bibnamefont {Sorokin}},\ }\bibfield  {title} {\bibinfo {title} {{Secondary laser cooling and capturing of thulium atoms in traps}},\ }\href {https://doi.org/10.1070/QE2014v044n06ABEH015392} {\bibfield  {journal} {\bibinfo  {journal} {Quantum Electron.}\ }\textbf {\bibinfo {volume} {44}},\ \bibinfo {pages} {515} (\bibinfo {year} {2014})}\BibitemShut {NoStop}%
\bibitem [{\citenamefont {Provorchenko}\ \emph {et~al.}(2023)\citenamefont {Provorchenko}, \citenamefont {Tregubov}, \citenamefont {Mishin}, \citenamefont {Yaushev}, \citenamefont {Kryuchkov}, \citenamefont {Sorokin}, \citenamefont {Khabarova}, \citenamefont {Golovizin},\ and\ \citenamefont {Kolachevsky}}]{Provorchenko.2023}%
  \BibitemOpen
  \bibfield  {author} {\bibinfo {author} {\bibfnamefont {D.}~\bibnamefont {Provorchenko}}, \bibinfo {author} {\bibfnamefont {D.}~\bibnamefont {Tregubov}}, \bibinfo {author} {\bibfnamefont {D.}~\bibnamefont {Mishin}}, \bibinfo {author} {\bibfnamefont {M.}~\bibnamefont {Yaushev}}, \bibinfo {author} {\bibfnamefont {D.}~\bibnamefont {Kryuchkov}}, \bibinfo {author} {\bibfnamefont {V.}~\bibnamefont {Sorokin}}, \bibinfo {author} {\bibfnamefont {K.}~\bibnamefont {Khabarova}}, \bibinfo {author} {\bibfnamefont {A.}~\bibnamefont {Golovizin}},\ and\ \bibinfo {author} {\bibfnamefont {N.}~\bibnamefont {Kolachevsky}},\ }\bibfield  {title} {\bibinfo {title} {{Deep Laser Cooling of Thulium Atoms to Sub-$\mu$K Temperatures in Magneto-Optical Trap}},\ }\bibfield  {journal} {\bibinfo  {journal} {Atoms}\ }\textbf {\bibinfo {volume} {11}},\ \href {https://doi.org/10.3390/atoms11020030} {10.3390/atoms11020030} (\bibinfo {year} {2023})\BibitemShut {NoStop}%
\bibitem [{\citenamefont {Mishin}\ \emph {et~al.}(2025)\citenamefont {Mishin}, \citenamefont {Tregubov}, \citenamefont {Kolachevsky},\ and\ \citenamefont {Golovizin}}]{Mishin.2025}%
  \BibitemOpen
  \bibfield  {author} {\bibinfo {author} {\bibfnamefont {D.}~\bibnamefont {Mishin}}, \bibinfo {author} {\bibfnamefont {D.}~\bibnamefont {Tregubov}}, \bibinfo {author} {\bibfnamefont {N.}~\bibnamefont {Kolachevsky}},\ and\ \bibinfo {author} {\bibfnamefont {A.}~\bibnamefont {Golovizin}},\ }\bibfield  {title} {\bibinfo {title} {Coherence of microwave and optical qubit levels in neutral thulium},\ }\href {https://doi.org/10.1103/f8xg-w57m} {\bibfield  {journal} {\bibinfo  {journal} {PRX Quantum}\ }\textbf {\bibinfo {volume} {6}},\ \bibinfo {pages} {040329} (\bibinfo {year} {2025})}\BibitemShut {NoStop}%
\bibitem [{\citenamefont {Sukachev}\ \emph {et~al.}(2016)\citenamefont {Sukachev}, \citenamefont {Fedorov}, \citenamefont {Tolstikhina}, \citenamefont {Tregubov}, \citenamefont {Kalganova}, \citenamefont {Vishnyakova}, \citenamefont {Golovizin}, \citenamefont {Kolachevsky}, \citenamefont {Khabarova},\ and\ \citenamefont {Sorokin}}]{Sukachev.2016}%
  \BibitemOpen
  \bibfield  {author} {\bibinfo {author} {\bibfnamefont {D.}~\bibnamefont {Sukachev}}, \bibinfo {author} {\bibfnamefont {S.}~\bibnamefont {Fedorov}}, \bibinfo {author} {\bibfnamefont {I.}~\bibnamefont {Tolstikhina}}, \bibinfo {author} {\bibfnamefont {D.}~\bibnamefont {Tregubov}}, \bibinfo {author} {\bibfnamefont {E.}~\bibnamefont {Kalganova}}, \bibinfo {author} {\bibfnamefont {G.}~\bibnamefont {Vishnyakova}}, \bibinfo {author} {\bibfnamefont {A.}~\bibnamefont {Golovizin}}, \bibinfo {author} {\bibfnamefont {N.}~\bibnamefont {Kolachevsky}}, \bibinfo {author} {\bibfnamefont {K.}~\bibnamefont {Khabarova}},\ and\ \bibinfo {author} {\bibfnamefont {V.}~\bibnamefont {Sorokin}},\ }\bibfield  {title} {\bibinfo {title} {{Inner-shell magnetic dipole transition in Tm atoms: A candidate for optical lattice clocks}},\ }\href {https://doi.org/10.1103/PhysRevA.94.022512} {\bibfield  {journal} {\bibinfo  {journal} {Phys. Rev. A}\ }\textbf {\bibinfo {volume} {94}},\ \bibinfo {pages} {022512} (\bibinfo {year}
  {2016})}\BibitemShut {NoStop}%
\bibitem [{\citenamefont {Golovizin}\ \emph {et~al.}(2019)\citenamefont {Golovizin}, \citenamefont {Fedorova}, \citenamefont {Tregubov}, \citenamefont {Sukachev}, \citenamefont {Khabarova}, \citenamefont {Sorokin},\ and\ \citenamefont {Kolachevsky}}]{Golovizin.2019}%
  \BibitemOpen
  \bibfield  {author} {\bibinfo {author} {\bibfnamefont {A.}~\bibnamefont {Golovizin}}, \bibinfo {author} {\bibfnamefont {E.}~\bibnamefont {Fedorova}}, \bibinfo {author} {\bibfnamefont {D.}~\bibnamefont {Tregubov}}, \bibinfo {author} {\bibfnamefont {D.}~\bibnamefont {Sukachev}}, \bibinfo {author} {\bibfnamefont {K.}~\bibnamefont {Khabarova}}, \bibinfo {author} {\bibfnamefont {V.}~\bibnamefont {Sorokin}},\ and\ \bibinfo {author} {\bibfnamefont {N.}~\bibnamefont {Kolachevsky}},\ }\bibfield  {title} {\bibinfo {title} {Inner-shell clock transition in atomic thulium with a small blackbody radiation shift},\ }\href {https://doi.org/10.1038/s41467-019-09706-9} {\bibfield  {journal} {\bibinfo  {journal} {Nat. Commun.}\ }\textbf {\bibinfo {volume} {10}},\ \bibinfo {pages} {1724} (\bibinfo {year} {2019})}\BibitemShut {NoStop}%
\bibitem [{\citenamefont {Golovizin}\ \emph {et~al.}(2024)\citenamefont {Golovizin}, \citenamefont {Mishin}, \citenamefont {Provorchenko}, \citenamefont {Tregubov},\ and\ \citenamefont {Kolachevsky}}]{Golovizin.2024}%
  \BibitemOpen
  \bibfield  {author} {\bibinfo {author} {\bibfnamefont {A.}~\bibnamefont {Golovizin}}, \bibinfo {author} {\bibfnamefont {D.}~\bibnamefont {Mishin}}, \bibinfo {author} {\bibfnamefont {D.}~\bibnamefont {Provorchenko}}, \bibinfo {author} {\bibfnamefont {D.}~\bibnamefont {Tregubov}},\ and\ \bibinfo {author} {\bibfnamefont {N.}~\bibnamefont {Kolachevsky}},\ }\bibfield  {title} {\bibinfo {title} {Synchronous {Comparison} of {Two} {Thulium} {Optical} {Clocks}},\ }\href {https://doi.org/10.1134/S0021364024600873} {\bibfield  {journal} {\bibinfo  {journal} {JETP Lett.}\ }\textbf {\bibinfo {volume} {119}},\ \bibinfo {pages} {659} (\bibinfo {year} {2024})}\BibitemShut {NoStop}%
\bibitem [{\citenamefont {Khlebnikov}\ \emph {et~al.}(2019)\citenamefont {Khlebnikov}, \citenamefont {Pershin}, \citenamefont {Tsyganok}, \citenamefont {Davletov}, \citenamefont {Cojocaru}, \citenamefont {Fedorova}, \citenamefont {Buchachenko},\ and\ \citenamefont {Akimov}}]{Khlebnikov.2019}%
  \BibitemOpen
  \bibfield  {author} {\bibinfo {author} {\bibfnamefont {V.~A.}\ \bibnamefont {Khlebnikov}}, \bibinfo {author} {\bibfnamefont {D.~A.}\ \bibnamefont {Pershin}}, \bibinfo {author} {\bibfnamefont {V.~V.}\ \bibnamefont {Tsyganok}}, \bibinfo {author} {\bibfnamefont {E.~T.}\ \bibnamefont {Davletov}}, \bibinfo {author} {\bibfnamefont {I.~S.}\ \bibnamefont {Cojocaru}}, \bibinfo {author} {\bibfnamefont {E.~S.}\ \bibnamefont {Fedorova}}, \bibinfo {author} {\bibfnamefont {A.~A.}\ \bibnamefont {Buchachenko}},\ and\ \bibinfo {author} {\bibfnamefont {A.~V.}\ \bibnamefont {Akimov}},\ }\bibfield  {title} {\bibinfo {title} {{Random to Chaotic Statistic Transformation in Low-Field Fano-Feshbach Resonances of Cold Thulium Atoms}},\ }\href {https://doi.org/10.1103/PhysRevLett.123.213402} {\bibfield  {journal} {\bibinfo  {journal} {Phys. Rev. Lett.}\ }\textbf {\bibinfo {volume} {123}},\ \bibinfo {pages} {213402} (\bibinfo {year} {2019})}\BibitemShut {NoStop}%
\bibitem [{\citenamefont {Davletov}\ \emph {et~al.}(2020)\citenamefont {Davletov}, \citenamefont {Tsyganok}, \citenamefont {Khlebnikov}, \citenamefont {Pershin}, \citenamefont {Shaykin},\ and\ \citenamefont {Akimov}}]{Davletov.2020}%
  \BibitemOpen
  \bibfield  {author} {\bibinfo {author} {\bibfnamefont {E.~T.}\ \bibnamefont {Davletov}}, \bibinfo {author} {\bibfnamefont {V.~V.}\ \bibnamefont {Tsyganok}}, \bibinfo {author} {\bibfnamefont {V.~A.}\ \bibnamefont {Khlebnikov}}, \bibinfo {author} {\bibfnamefont {D.~A.}\ \bibnamefont {Pershin}}, \bibinfo {author} {\bibfnamefont {D.~V.}\ \bibnamefont {Shaykin}},\ and\ \bibinfo {author} {\bibfnamefont {A.~V.}\ \bibnamefont {Akimov}},\ }\bibfield  {title} {\bibinfo {title} {{Machine learning for achieving Bose-Einstein condensation of thulium atoms}},\ }\href {https://doi.org/10.1103/PhysRevA.102.011302} {\bibfield  {journal} {\bibinfo  {journal} {Phys. Rev. A}\ }\textbf {\bibinfo {volume} {102}},\ \bibinfo {pages} {011302(R)} (\bibinfo {year} {2020})}\BibitemShut {NoStop}%
\bibitem [{\citenamefont {Tsyganok}\ \emph {et~al.}(2023)\citenamefont {Tsyganok}, \citenamefont {Pershin}, \citenamefont {Khlebnikov}, \citenamefont {Kumpilov}, \citenamefont {Pyrkh}, \citenamefont {Rudnev}, \citenamefont {Fedotova}, \citenamefont {Gaifudinov}, \citenamefont {Cojocaru}, \citenamefont {Khoruzhii}, \citenamefont {Aksentsev}, \citenamefont {Zykova},\ and\ \citenamefont {Akimov}}]{Tsyganok.2023}%
  \BibitemOpen
  \bibfield  {author} {\bibinfo {author} {\bibfnamefont {V.~V.}\ \bibnamefont {Tsyganok}}, \bibinfo {author} {\bibfnamefont {D.~A.}\ \bibnamefont {Pershin}}, \bibinfo {author} {\bibfnamefont {V.~A.}\ \bibnamefont {Khlebnikov}}, \bibinfo {author} {\bibfnamefont {D.~A.}\ \bibnamefont {Kumpilov}}, \bibinfo {author} {\bibfnamefont {I.~A.}\ \bibnamefont {Pyrkh}}, \bibinfo {author} {\bibfnamefont {A.~E.}\ \bibnamefont {Rudnev}}, \bibinfo {author} {\bibfnamefont {E.~A.}\ \bibnamefont {Fedotova}}, \bibinfo {author} {\bibfnamefont {D.~V.}\ \bibnamefont {Gaifudinov}}, \bibinfo {author} {\bibfnamefont {I.~S.}\ \bibnamefont {Cojocaru}}, \bibinfo {author} {\bibfnamefont {K.~A.}\ \bibnamefont {Khoruzhii}}, \bibinfo {author} {\bibfnamefont {P.~A.}\ \bibnamefont {Aksentsev}}, \bibinfo {author} {\bibfnamefont {A.~K.}\ \bibnamefont {Zykova}},\ and\ \bibinfo {author} {\bibfnamefont {A.~V.}\ \bibnamefont {Akimov}},\ }\bibfield  {title} {\bibinfo {title} {{Bose-Einstein} condensate as a diagnostic tool for an optical lattice
  formed by 1064-nm laser light},\ }\href {https://doi.org/10.1103/PhysRevA.108.013310} {\bibfield  {journal} {\bibinfo  {journal} {Phys. Rev. A}\ }\textbf {\bibinfo {volume} {108}},\ \bibinfo {pages} {013310} (\bibinfo {year} {2023})}\BibitemShut {NoStop}%
\bibitem [{\citenamefont {Allcock}\ \emph {et~al.}(2021)\citenamefont {Allcock}, \citenamefont {Campbell}, \citenamefont {Chiaverini}, \citenamefont {Chuang}, \citenamefont {Hudson}, \citenamefont {Moore}, \citenamefont {Ransford}, \citenamefont {Roman}, \citenamefont {Sage},\ and\ \citenamefont {Wineland}}]{Allcock.2021}%
  \BibitemOpen
  \bibfield  {author} {\bibinfo {author} {\bibfnamefont {D.~T.~C.}\ \bibnamefont {Allcock}}, \bibinfo {author} {\bibfnamefont {W.~C.}\ \bibnamefont {Campbell}}, \bibinfo {author} {\bibfnamefont {J.}~\bibnamefont {Chiaverini}}, \bibinfo {author} {\bibfnamefont {I.~L.}\ \bibnamefont {Chuang}}, \bibinfo {author} {\bibfnamefont {E.~R.}\ \bibnamefont {Hudson}}, \bibinfo {author} {\bibfnamefont {I.~D.}\ \bibnamefont {Moore}}, \bibinfo {author} {\bibfnamefont {A.}~\bibnamefont {Ransford}}, \bibinfo {author} {\bibfnamefont {C.}~\bibnamefont {Roman}}, \bibinfo {author} {\bibfnamefont {J.~M.}\ \bibnamefont {Sage}},\ and\ \bibinfo {author} {\bibfnamefont {D.~J.}\ \bibnamefont {Wineland}},\ }\bibfield  {title} {\bibinfo {title} {omg blueprint for trapped ion quantum computing with metastable states},\ }\href {https://doi.org/10.1063/5.0069544} {\bibfield  {journal} {\bibinfo  {journal} {Appl. Phys. Lett.}\ }\textbf {\bibinfo {volume} {119}},\ \bibinfo {pages} {214002} (\bibinfo {year} {2021})}\BibitemShut {NoStop}%
\bibitem [{\citenamefont {Bi{\'e}mont}\ \emph {et~al.}(1999)\citenamefont {Bi{\'e}mont}, \citenamefont {Palmeri},\ and\ \citenamefont {Quinet}}]{Biemont.1999}%
  \BibitemOpen
  \bibfield  {author} {\bibinfo {author} {\bibfnamefont {{\'E}.}~\bibnamefont {Bi{\'e}mont}}, \bibinfo {author} {\bibfnamefont {P.}~\bibnamefont {Palmeri}},\ and\ \bibinfo {author} {\bibfnamefont {P.}~\bibnamefont {Quinet}},\ }\bibfield  {title} {\bibinfo {title} {{A New Database of Astrophysical Interest}},\ }\href {https://doi.org/10.1023/A:1017049314691} {\bibfield  {journal} {\bibinfo  {journal} {Astrophys. Space Sci.}\ }\textbf {\bibinfo {volume} {269}},\ \bibinfo {pages} {635} (\bibinfo {year} {1999})},\ \bibinfo {note} {\url{https://agif.umons.ac.be/databases/dream.html}}\BibitemShut {NoStop}%
\bibitem [{\citenamefont {Jain}\ \emph {et~al.}(2024)\citenamefont {Jain}, \citenamefont {Hudson}, \citenamefont {Campbell},\ and\ \citenamefont {Albert}}]{Shubham.2024}%
  \BibitemOpen
  \bibfield  {author} {\bibinfo {author} {\bibfnamefont {S.~P.}\ \bibnamefont {Jain}}, \bibinfo {author} {\bibfnamefont {E.~R.}\ \bibnamefont {Hudson}}, \bibinfo {author} {\bibfnamefont {W.~C.}\ \bibnamefont {Campbell}},\ and\ \bibinfo {author} {\bibfnamefont {V.~V.}\ \bibnamefont {Albert}},\ }\bibfield  {title} {\bibinfo {title} {Absorption-emission codes for atomic and molecular quantum information platforms},\ }\href {https://doi.org/10.1103/PhysRevLett.133.260601} {\bibfield  {journal} {\bibinfo  {journal} {Phys. Rev. Lett.}\ }\textbf {\bibinfo {volume} {133}},\ \bibinfo {pages} {260601} (\bibinfo {year} {2024})}\BibitemShut {NoStop}%
\bibitem [{\citenamefont {Aydin}\ and\ \citenamefont {Barg}(2025)}]{Aydin.2025}%
  \BibitemOpen
  \bibfield  {author} {\bibinfo {author} {\bibfnamefont {A.}~\bibnamefont {Aydin}}\ and\ \bibinfo {author} {\bibfnamefont {A.}~\bibnamefont {Barg}},\ }\bibfield  {title} {\bibinfo {title} {Class of codes correcting absorptions and emissions},\ }\href {https://doi.org/10.1103/PhysRevA.111.022415} {\bibfield  {journal} {\bibinfo  {journal} {Phys. Rev. A}\ }\textbf {\bibinfo {volume} {111}},\ \bibinfo {pages} {022415} (\bibinfo {year} {2025})}\BibitemShut {NoStop}%
\bibitem [{\citenamefont {Lepers}\ \emph {et~al.}(2016)\citenamefont {Lepers}, \citenamefont {Hong}, \citenamefont {Wyart},\ and\ \citenamefont {Dulieu}}]{Lepers.2016}%
  \BibitemOpen
  \bibfield  {author} {\bibinfo {author} {\bibfnamefont {M.}~\bibnamefont {Lepers}}, \bibinfo {author} {\bibfnamefont {Y.}~\bibnamefont {Hong}}, \bibinfo {author} {\bibfnamefont {J.-F.}\ \bibnamefont {Wyart}},\ and\ \bibinfo {author} {\bibfnamefont {O.}~\bibnamefont {Dulieu}},\ }\bibfield  {title} {\bibinfo {title} {Proposal for laser cooling of rare-earth ions},\ }\href {https://doi.org/10.1103/PhysRevA.93.011401} {\bibfield  {journal} {\bibinfo  {journal} {Phys. Rev. A}\ }\textbf {\bibinfo {volume} {93}},\ \bibinfo {pages} {011401(R)} (\bibinfo {year} {2016})}\BibitemShut {NoStop}%
\bibitem [{\citenamefont {M{\"u}ller}\ \emph {et~al.}(2026)\citenamefont {M{\"u}ller}, \citenamefont {Tretiakov}, \citenamefont {Younes}, \citenamefont {Halawani}, \citenamefont {C.~Campbell},\ and\ \citenamefont {Hamilton}}]{Mueller.2026b}%
  \BibitemOpen
  \bibfield  {author} {\bibinfo {author} {\bibfnamefont {P.}~\bibnamefont {M{\"u}ller}}, \bibinfo {author} {\bibfnamefont {A.}~\bibnamefont {Tretiakov}}, \bibinfo {author} {\bibfnamefont {A.}~\bibnamefont {Younes}}, \bibinfo {author} {\bibfnamefont {N.}~\bibnamefont {Halawani}}, \bibinfo {author} {\bibfnamefont {W.}~\bibnamefont {C.~Campbell}},\ and\ \bibinfo {author} {\bibfnamefont {P.}~\bibnamefont {Hamilton}},\ }\bibfield  {title} {\bibinfo {title} {Hyperfine spectroscopy of optical-cycling transitions in singly ionized thulium},\ }\href {https://doi.org/10.1038/s41598-026-45288-5} {\bibfield  {journal} {\bibinfo  {journal} {Scient. Rep.}\ }\textbf {\bibinfo {volume} {16}},\ \bibinfo {pages} {10655} (\bibinfo {year} {2026})}\BibitemShut {NoStop}%
\bibitem [{\citenamefont {Cheal}\ and\ \citenamefont {Flanagan}(2010)}]{Cheal.2010}%
  \BibitemOpen
  \bibfield  {author} {\bibinfo {author} {\bibfnamefont {B.}~\bibnamefont {Cheal}}\ and\ \bibinfo {author} {\bibfnamefont {K.~T.}\ \bibnamefont {Flanagan}},\ }\bibfield  {title} {\bibinfo {title} {Progress in laser spectroscopy at radioactive ion beam facilities},\ }\href {https://doi.org/10.1088/0954-3899/37/11/113101} {\bibfield  {journal} {\bibinfo  {journal} {J. Phys. G}\ }\textbf {\bibinfo {volume} {37}},\ \bibinfo {pages} {113101} (\bibinfo {year} {2010})}\BibitemShut {NoStop}%
\bibitem [{\citenamefont {Blaum}\ \emph {et~al.}(2013)\citenamefont {Blaum}, \citenamefont {Dilling},\ and\ \citenamefont {Nörtershäuser}}]{Blaum.2013}%
  \BibitemOpen
  \bibfield  {author} {\bibinfo {author} {\bibfnamefont {K.}~\bibnamefont {Blaum}}, \bibinfo {author} {\bibfnamefont {J.}~\bibnamefont {Dilling}},\ and\ \bibinfo {author} {\bibfnamefont {W.}~\bibnamefont {Nörtershäuser}},\ }\bibfield  {title} {\bibinfo {title} {Precision {Atomic} {Physics} {Techniques} for {Nuclear} {Physics} with {Radioactive} {Beams}},\ }\href {https://doi.org/10.1088/0031-8949/2013/T152/014017} {\bibfield  {journal} {\bibinfo  {journal} {Phys. Scr.}\ }\textbf {\bibinfo {volume} {T152}},\ \bibinfo {pages} {014017} (\bibinfo {year} {2013})}\BibitemShut {NoStop}%
\bibitem [{\citenamefont {Campbell}\ \emph {et~al.}(2016)\citenamefont {Campbell}, \citenamefont {Moore},\ and\ \citenamefont {Pearson}}]{Campbell.2016}%
  \BibitemOpen
  \bibfield  {author} {\bibinfo {author} {\bibfnamefont {P.}~\bibnamefont {Campbell}}, \bibinfo {author} {\bibfnamefont {I.~D.}\ \bibnamefont {Moore}},\ and\ \bibinfo {author} {\bibfnamefont {M.~R.}\ \bibnamefont {Pearson}},\ }\bibfield  {title} {\bibinfo {title} {Laser spectroscopy for nuclear structure physics},\ }\href {https://doi.org/10.1016/j.ppnp.2015.09.003} {\bibfield  {journal} {\bibinfo  {journal} {Progr. Part. Nucl. Phys.}\ }\textbf {\bibinfo {volume} {86}},\ \bibinfo {pages} {127} (\bibinfo {year} {2016})}\BibitemShut {NoStop}%
\bibitem [{\citenamefont {Neugart}\ \emph {et~al.}(2017)\citenamefont {Neugart}, \citenamefont {Billowes}, \citenamefont {Bissell}, \citenamefont {Blaum}, \citenamefont {Cheal}, \citenamefont {Flanagan}, \citenamefont {Neyens}, \citenamefont {Nörtershäuser},\ and\ \citenamefont {Yordanov}}]{Neugart.2017}%
  \BibitemOpen
  \bibfield  {author} {\bibinfo {author} {\bibfnamefont {R.}~\bibnamefont {Neugart}}, \bibinfo {author} {\bibfnamefont {J.}~\bibnamefont {Billowes}}, \bibinfo {author} {\bibfnamefont {M.~L.}\ \bibnamefont {Bissell}}, \bibinfo {author} {\bibfnamefont {K.}~\bibnamefont {Blaum}}, \bibinfo {author} {\bibfnamefont {B.}~\bibnamefont {Cheal}}, \bibinfo {author} {\bibfnamefont {K.~T.}\ \bibnamefont {Flanagan}}, \bibinfo {author} {\bibfnamefont {G.}~\bibnamefont {Neyens}}, \bibinfo {author} {\bibfnamefont {W.}~\bibnamefont {Nörtershäuser}},\ and\ \bibinfo {author} {\bibfnamefont {D.~T.}\ \bibnamefont {Yordanov}},\ }\bibfield  {title} {\bibinfo {title} {Collinear laser spectroscopy at {ISOLDE}: new methods and highlights},\ }\href {https://doi.org/10.1088/1361-6471/aa6642} {\bibfield  {journal} {\bibinfo  {journal} {J. Phys. G}\ }\textbf {\bibinfo {volume} {44}},\ \bibinfo {pages} {064002} (\bibinfo {year} {2017})}\BibitemShut {NoStop}%
\bibitem [{\citenamefont {Yang}\ \emph {et~al.}(2023)\citenamefont {Yang}, \citenamefont {Wang}, \citenamefont {Wilkins},\ and\ \citenamefont {Ruiz}}]{Yang.2023}%
  \BibitemOpen
  \bibfield  {author} {\bibinfo {author} {\bibfnamefont {X.~F.}\ \bibnamefont {Yang}}, \bibinfo {author} {\bibfnamefont {S.~J.}\ \bibnamefont {Wang}}, \bibinfo {author} {\bibfnamefont {S.~G.}\ \bibnamefont {Wilkins}},\ and\ \bibinfo {author} {\bibfnamefont {R.~G.}\ \bibnamefont {Ruiz}},\ }\bibfield  {title} {\bibinfo {title} {Laser spectroscopy for the study of exotic nuclei},\ }\href {https://doi.org/10.1016/j.ppnp.2022.104005} {\bibfield  {journal} {\bibinfo  {journal} {Progr. Part. Nucl. Phys.}\ }\textbf {\bibinfo {volume} {129}},\ \bibinfo {pages} {104005} (\bibinfo {year} {2023})}\BibitemShut {NoStop}%
\bibitem [{\citenamefont {Alkhazov}\ \emph {et~al.}(1988)\citenamefont {Alkhazov}, \citenamefont {Barzakh}, \citenamefont {Chubukov}, \citenamefont {Denisov}, \citenamefont {Ivanov}, \citenamefont {Panteleev}, \citenamefont {Starodubsky}, \citenamefont {Buyanov}, \citenamefont {Fedoseyev}, \citenamefont {Letokhov}, \citenamefont {Mishin},\ and\ \citenamefont {Sekatskli}}]{Alkhazov.1988}%
  \BibitemOpen
  \bibfield  {author} {\bibinfo {author} {\bibfnamefont {G.~D.}\ \bibnamefont {Alkhazov}}, \bibinfo {author} {\bibfnamefont {A.~E.}\ \bibnamefont {Barzakh}}, \bibinfo {author} {\bibfnamefont {I.}~\bibnamefont {Chubukov}}, \bibinfo {author} {\bibfnamefont {V.~P.}\ \bibnamefont {Denisov}}, \bibinfo {author} {\bibfnamefont {V.~S.}\ \bibnamefont {Ivanov}}, \bibinfo {author} {\bibfnamefont {V.~N.}\ \bibnamefont {Panteleev}}, \bibinfo {author} {\bibfnamefont {V.~E.}\ \bibnamefont {Starodubsky}}, \bibinfo {author} {\bibfnamefont {N.~B.}\ \bibnamefont {Buyanov}}, \bibinfo {author} {\bibfnamefont {M.~N.}\ \bibnamefont {Fedoseyev}}, \bibinfo {author} {\bibfnamefont {V.~S.}\ \bibnamefont {Letokhov}}, \bibinfo {author} {\bibfnamefont {V.~I.}\ \bibnamefont {Mishin}},\ and\ \bibinfo {author} {\bibfnamefont {S.~K.}\ \bibnamefont {Sekatskli}},\ }\bibfield  {title} {\bibinfo {title} {Nuclear electromagnetic moments and charge radii of deformed thulium isotopes with the mass numbers {A} = 157--172},\ }\href
  {https://doi.org/10.1016/0375-9474(88)90359-4} {\bibfield  {journal} {\bibinfo  {journal} {Nucl. Phys. A}\ }\textbf {\bibinfo {volume} {477}},\ \bibinfo {pages} {37} (\bibinfo {year} {1988})}\BibitemShut {NoStop}%
\bibitem [{\citenamefont {Barzakh}\ \emph {et~al.}(2000)\citenamefont {Barzakh}, \citenamefont {Chubukov}, \citenamefont {Fedorov}, \citenamefont {Panteleev}, \citenamefont {Seliverstov},\ and\ \citenamefont {Volkov}}]{Barzakh.2000}%
  \BibitemOpen
  \bibfield  {author} {\bibinfo {author} {\bibfnamefont {A.~E.}\ \bibnamefont {Barzakh}}, \bibinfo {author} {\bibfnamefont {I.~Y.}\ \bibnamefont {Chubukov}}, \bibinfo {author} {\bibfnamefont {D.~V.}\ \bibnamefont {Fedorov}}, \bibinfo {author} {\bibfnamefont {V.~N.}\ \bibnamefont {Panteleev}}, \bibinfo {author} {\bibfnamefont {M.~D.}\ \bibnamefont {Seliverstov}},\ and\ \bibinfo {author} {\bibfnamefont {Y.~M.}\ \bibnamefont {Volkov}},\ }\bibfield  {title} {\bibinfo {title} {{Mean square charge radii of the neutron-deficient rare-earth isotopes in the region of the nuclear shell $N=82$ measured by the laser ion source spectroscopy technique}},\ }\href {https://doi.org/10.1103/PhysRevC.61.034304} {\bibfield  {journal} {\bibinfo  {journal} {Phys. Rev. C}\ }\textbf {\bibinfo {volume} {61}},\ \bibinfo {pages} {034304} (\bibinfo {year} {2000})}\BibitemShut {NoStop}%
\bibitem [{\citenamefont {Weyrich}\ \emph {et~al.}(2026)\citenamefont {Weyrich}, \citenamefont {van Beek}, \citenamefont {XXX}, \citenamefont {Arya}, \citenamefont {Berndt}, \citenamefont {Block}, \citenamefont {Brizard}, \citenamefont {Chhetri}, \citenamefont {Claessens}, \citenamefont {Düllmann}, \citenamefont {Ferrer}, \citenamefont {Geldhof}, \citenamefont {Giacoppo}, \citenamefont {Gutierrez}, \citenamefont {Hasse}, \citenamefont {Helmel}, \citenamefont {Heßberger}, \citenamefont {Hindermann}, \citenamefont {Ivandikov}, \citenamefont {Jana}, \citenamefont {Kieck}, \citenamefont {Laatiaoui}, \citenamefont {Lecesne}, \citenamefont {Mistry}, \citenamefont {Münzberg}, \citenamefont {Niemeyer}, \citenamefont {Raeder}, \citenamefont {Rickert}, \citenamefont {Rodriguez}, \citenamefont {Savajols}, \citenamefont {Stemmler}, \citenamefont {Studer}, \citenamefont {Walther}, \citenamefont {Warbinek}, \citenamefont {Duppen},\ and\ \citenamefont {Wendt}}]{Weyrich.2026}%
  \BibitemOpen
  \bibfield  {author} {\bibinfo {author} {\bibfnamefont {J.}~\bibnamefont {Weyrich}}, \bibinfo {author} {\bibfnamefont {K.}~\bibnamefont {van Beek}}, \bibinfo {author} {\bibfnamefont {H.}~\bibnamefont {XXX}}, \bibinfo {author} {\bibfnamefont {A.}~\bibnamefont {Arya}}, \bibinfo {author} {\bibfnamefont {S.}~\bibnamefont {Berndt}}, \bibinfo {author} {\bibfnamefont {M.}~\bibnamefont {Block}}, \bibinfo {author} {\bibfnamefont {A.}~\bibnamefont {Brizard}}, \bibinfo {author} {\bibfnamefont {P.}~\bibnamefont {Chhetri}}, \bibinfo {author} {\bibfnamefont {A.}~\bibnamefont {Claessens}}, \bibinfo {author} {\bibfnamefont {C.~E.}\ \bibnamefont {Düllmann}}, \bibinfo {author} {\bibfnamefont {R.}~\bibnamefont {Ferrer}}, \bibinfo {author} {\bibfnamefont {S.}~\bibnamefont {Geldhof}}, \bibinfo {author} {\bibfnamefont {F.}~\bibnamefont {Giacoppo}}, \bibinfo {author} {\bibfnamefont {M.~J.}\ \bibnamefont {Gutierrez}}, \bibinfo {author} {\bibfnamefont {R.}~\bibnamefont {Hasse}}, \bibinfo {author} {\bibfnamefont {C.}~\bibnamefont
  {Helmel}}, \bibinfo {author} {\bibfnamefont {F.~P.}\ \bibnamefont {Heßberger}}, \bibinfo {author} {\bibfnamefont {J.}~\bibnamefont {Hindermann}}, \bibinfo {author} {\bibfnamefont {F.}~\bibnamefont {Ivandikov}}, \bibinfo {author} {\bibfnamefont {B.}~\bibnamefont {Jana}}, \bibinfo {author} {\bibfnamefont {T.}~\bibnamefont {Kieck}}, \bibinfo {author} {\bibfnamefont {M.}~\bibnamefont {Laatiaoui}}, \bibinfo {author} {\bibfnamefont {N.}~\bibnamefont {Lecesne}}, \bibinfo {author} {\bibfnamefont {A.}~\bibnamefont {Mistry}}, \bibinfo {author} {\bibfnamefont {D.}~\bibnamefont {Münzberg}}, \bibinfo {author} {\bibfnamefont {T.}~\bibnamefont {Niemeyer}}, \bibinfo {author} {\bibfnamefont {S.}~\bibnamefont {Raeder}}, \bibinfo {author} {\bibfnamefont {E.}~\bibnamefont {Rickert}}, \bibinfo {author} {\bibfnamefont {D.}~\bibnamefont {Rodriguez}}, \bibinfo {author} {\bibfnamefont {H.}~\bibnamefont {Savajols}}, \bibinfo {author} {\bibfnamefont {M.}~\bibnamefont {Stemmler}}, \bibinfo {author} {\bibfnamefont {D.}~\bibnamefont
  {Studer}}, \bibinfo {author} {\bibfnamefont {T.}~\bibnamefont {Walther}}, \bibinfo {author} {\bibfnamefont {J.}~\bibnamefont {Warbinek}}, \bibinfo {author} {\bibfnamefont {P.~V.}\ \bibnamefont {Duppen}},\ and\ \bibinfo {author} {\bibfnamefont {K.}~\bibnamefont {Wendt}},\ }\href {https://arxiv.org/abs/2607.25033} {\bibinfo {title} {{Laser Spectroscopy of Thulium Isotopes Near the ($N$=82) Shell Closure: Nuclear Moment and Charge Radius of ${}^{152\mathrm{m}}\mathrm{Tm}$}}} (\bibinfo {year} {2026}),\ \Eprint {https://arxiv.org/abs/2607.25033} {arXiv:2607.25033 [nucl-ex]} \BibitemShut {NoStop}%
\bibitem [{\citenamefont {Cheal}\ \emph {et~al.}(2022)\citenamefont {Cheal}, \citenamefont {Rodríguez}, \citenamefont {Bai}, \citenamefont {Blaum}, \citenamefont {Campbell}, \citenamefont {Garcia~Ruiz}, \citenamefont {Imgram}, \citenamefont {König}, \citenamefont {Lellinger}, \citenamefont {Müller}, \citenamefont {Nazarewicz}, \citenamefont {Neugart}, \citenamefont {Neyens}, \citenamefont {Nies}, \citenamefont {Nörtershäuser}, \citenamefont {Page}, \citenamefont {Plattner}, \citenamefont {Reinhard}, \citenamefont {Renth}, \citenamefont {Rothe}, \citenamefont {Sánchez}, \citenamefont {Schweiger}, \citenamefont {Stegemann}, \citenamefont {Stora}, \citenamefont {Wang}, \citenamefont {Yang},\ and\ \citenamefont {Yordanov}}]{Cheal_ISOLDE_LoI}%
  \BibitemOpen
  \bibfield  {author} {\bibinfo {author} {\bibfnamefont {B.}~\bibnamefont {Cheal}}, \bibinfo {author} {\bibfnamefont {L.}~\bibnamefont {Rodríguez}}, \bibinfo {author} {\bibfnamefont {S.}~\bibnamefont {Bai}}, \bibinfo {author} {\bibfnamefont {K.}~\bibnamefont {Blaum}}, \bibinfo {author} {\bibfnamefont {P.}~\bibnamefont {Campbell}}, \bibinfo {author} {\bibfnamefont {R.}~\bibnamefont {Garcia~Ruiz}}, \bibinfo {author} {\bibfnamefont {P.}~\bibnamefont {Imgram}}, \bibinfo {author} {\bibfnamefont {K.}~\bibnamefont {König}}, \bibinfo {author} {\bibfnamefont {T.}~\bibnamefont {Lellinger}}, \bibinfo {author} {\bibfnamefont {P.}~\bibnamefont {Müller}}, \bibinfo {author} {\bibfnamefont {W.}~\bibnamefont {Nazarewicz}}, \bibinfo {author} {\bibfnamefont {R.}~\bibnamefont {Neugart}}, \bibinfo {author} {\bibfnamefont {G.}~\bibnamefont {Neyens}}, \bibinfo {author} {\bibfnamefont {L.}~\bibnamefont {Nies}}, \bibinfo {author} {\bibfnamefont {W.}~\bibnamefont {Nörtershäuser}}, \bibinfo {author} {\bibfnamefont
  {R.}~\bibnamefont {Page}}, \bibinfo {author} {\bibfnamefont {P.}~\bibnamefont {Plattner}}, \bibinfo {author} {\bibfnamefont {P.-G.}\ \bibnamefont {Reinhard}}, \bibinfo {author} {\bibfnamefont {L.}~\bibnamefont {Renth}}, \bibinfo {author} {\bibfnamefont {S.}~\bibnamefont {Rothe}}, \bibinfo {author} {\bibfnamefont {R.}~\bibnamefont {Sánchez}}, \bibinfo {author} {\bibfnamefont {C.}~\bibnamefont {Schweiger}}, \bibinfo {author} {\bibfnamefont {S.}~\bibnamefont {Stegemann}}, \bibinfo {author} {\bibfnamefont {T.}~\bibnamefont {Stora}}, \bibinfo {author} {\bibfnamefont {S.}~\bibnamefont {Wang}}, \bibinfo {author} {\bibfnamefont {X.}~\bibnamefont {Yang}},\ and\ \bibinfo {author} {\bibfnamefont {D.}~\bibnamefont {Yordanov}},\ }\href {https://cds.cern.ch/record/2834596} {\emph {\bibinfo {title} {{Letter of Intent INTC-I-245: Laser spectroscopy of neutron-deficient thulium isotopes}}}},\ \bibinfo {type} {Tech. Rep.}\ (\bibinfo  {institution} {CERN},\ \bibinfo {address} {Geneva},\ \bibinfo {year} {2022})\BibitemShut
  {NoStop}%
\bibitem [{\citenamefont {Cheal}\ \emph {et~al.}(2023)\citenamefont {Cheal}, \citenamefont {Rodríguez}, \citenamefont {Bai}, \citenamefont {Blaum}, \citenamefont {Campbell}, \citenamefont {Chrysalidis}, \citenamefont {Garcia~Ruiz}, \citenamefont {Giesel}, \citenamefont {Heinke}, \citenamefont {Imgram}, \citenamefont {König}, \citenamefont {Lange}, \citenamefont {Lellinger}, \citenamefont {Lunney}, \citenamefont {Lynch}, \citenamefont {Marsh}, \citenamefont {Müller}, \citenamefont {Nazarewicz}, \citenamefont {Neugart}, \citenamefont {Neyens}, \citenamefont {Nies}, \citenamefont {Nörtershäuser}, \citenamefont {Page}, \citenamefont {Plattner}, \citenamefont {Reinhard}, \citenamefont {Renth}, \citenamefont {Rothe}, \citenamefont {Sánchez}, \citenamefont {Schweiger}, \citenamefont {Schweikhard}, \citenamefont {Stegemann}, \citenamefont {Stora}, \citenamefont {Wang}, \citenamefont {Yang},\ and\ \citenamefont {Yordanov}}]{Cheal_ISOLDE_Proposal}%
  \BibitemOpen
  \bibfield  {author} {\bibinfo {author} {\bibfnamefont {B.}~\bibnamefont {Cheal}}, \bibinfo {author} {\bibfnamefont {L.}~\bibnamefont {Rodríguez}}, \bibinfo {author} {\bibfnamefont {S.}~\bibnamefont {Bai}}, \bibinfo {author} {\bibfnamefont {K.}~\bibnamefont {Blaum}}, \bibinfo {author} {\bibfnamefont {P.}~\bibnamefont {Campbell}}, \bibinfo {author} {\bibfnamefont {K.}~\bibnamefont {Chrysalidis}}, \bibinfo {author} {\bibfnamefont {R.}~\bibnamefont {Garcia~Ruiz}}, \bibinfo {author} {\bibfnamefont {P.}~\bibnamefont {Giesel}}, \bibinfo {author} {\bibfnamefont {R.}~\bibnamefont {Heinke}}, \bibinfo {author} {\bibfnamefont {P.}~\bibnamefont {Imgram}}, \bibinfo {author} {\bibfnamefont {K.}~\bibnamefont {König}}, \bibinfo {author} {\bibfnamefont {D.}~\bibnamefont {Lange}}, \bibinfo {author} {\bibfnamefont {T.}~\bibnamefont {Lellinger}}, \bibinfo {author} {\bibfnamefont {D.}~\bibnamefont {Lunney}}, \bibinfo {author} {\bibfnamefont {K.}~\bibnamefont {Lynch}}, \bibinfo {author} {\bibfnamefont {B.}~\bibnamefont
  {Marsh}}, \bibinfo {author} {\bibfnamefont {P.}~\bibnamefont {Müller}}, \bibinfo {author} {\bibfnamefont {W.}~\bibnamefont {Nazarewicz}}, \bibinfo {author} {\bibfnamefont {R.}~\bibnamefont {Neugart}}, \bibinfo {author} {\bibfnamefont {G.}~\bibnamefont {Neyens}}, \bibinfo {author} {\bibfnamefont {L.}~\bibnamefont {Nies}}, \bibinfo {author} {\bibfnamefont {W.}~\bibnamefont {Nörtershäuser}}, \bibinfo {author} {\bibfnamefont {R.}~\bibnamefont {Page}}, \bibinfo {author} {\bibfnamefont {P.}~\bibnamefont {Plattner}}, \bibinfo {author} {\bibfnamefont {P.-G.}\ \bibnamefont {Reinhard}}, \bibinfo {author} {\bibfnamefont {L.}~\bibnamefont {Renth}}, \bibinfo {author} {\bibfnamefont {S.}~\bibnamefont {Rothe}}, \bibinfo {author} {\bibfnamefont {R.}~\bibnamefont {Sánchez}}, \bibinfo {author} {\bibfnamefont {C.}~\bibnamefont {Schweiger}}, \bibinfo {author} {\bibfnamefont {L.}~\bibnamefont {Schweikhard}}, \bibinfo {author} {\bibfnamefont {S.}~\bibnamefont {Stegemann}}, \bibinfo {author} {\bibfnamefont {T.}~\bibnamefont
  {Stora}}, \bibinfo {author} {\bibfnamefont {S.}~\bibnamefont {Wang}}, \bibinfo {author} {\bibfnamefont {X.}~\bibnamefont {Yang}},\ and\ \bibinfo {author} {\bibfnamefont {D.}~\bibnamefont {Yordanov}},\ }\href {https://cds.cern.ch/record/2872390} {\emph {\bibinfo {title} {{Proposal INTC-P-673: Laser spectroscopy of neutron-deficient thulium isotopes}}}},\ \bibinfo {type} {Tech. Rep.}\ (\bibinfo  {institution} {CERN},\ \bibinfo {address} {Geneva},\ \bibinfo {year} {2023})\BibitemShut {NoStop}%
\bibitem [{\citenamefont {Wang}\ \emph {et~al.}(2022)\citenamefont {Wang}, \citenamefont {Yu}, \citenamefont {Tian}, \citenamefont {Chen}, \citenamefont {Xie}, \citenamefont {Zeng}, \citenamefont {Guo},\ and\ \citenamefont {Chang}}]{Wang.2022}%
  \BibitemOpen
  \bibfield  {author} {\bibinfo {author} {\bibfnamefont {X.}~\bibnamefont {Wang}}, \bibinfo {author} {\bibfnamefont {Q.}~\bibnamefont {Yu}}, \bibinfo {author} {\bibfnamefont {Y.}~\bibnamefont {Tian}}, \bibinfo {author} {\bibfnamefont {Z.}~\bibnamefont {Chen}}, \bibinfo {author} {\bibfnamefont {H.}~\bibnamefont {Xie}}, \bibinfo {author} {\bibfnamefont {X.}~\bibnamefont {Zeng}}, \bibinfo {author} {\bibfnamefont {G.}~\bibnamefont {Guo}},\ and\ \bibinfo {author} {\bibfnamefont {H.}~\bibnamefont {Chang}},\ }\bibfield  {title} {\bibinfo {title} {{Experimental branching fractions, transition probabilities and oscillator strengths in Tm I and Tm II}},\ }\href {https://doi.org/https://doi.org/10.1016/j.jqsrt.2022.108091} {\bibfield  {journal} {\bibinfo  {journal} {J. Quant. Spectrosc. Radiat. Transf.}\ }\textbf {\bibinfo {volume} {280}},\ \bibinfo {pages} {108091} (\bibinfo {year} {2022})}\BibitemShut {NoStop}%
\bibitem [{\citenamefont {{Cowley}}\ and\ \citenamefont {{Mathys}}(1998)}]{Cowley1998}%
  \BibitemOpen
  \bibfield  {author} {\bibinfo {author} {\bibfnamefont {C.~R.}\ \bibnamefont {{Cowley}}}\ and\ \bibinfo {author} {\bibfnamefont {G.}~\bibnamefont {{Mathys}}},\ }\bibfield  {title} {\bibinfo {title} {{{Line identifications and preliminary abundances from the red spectrum of HD 101065 (Przybylski’s star)}}},\ }\href {https://ui.adsabs.harvard.edu/abs/1998A&A...339..165C} {\bibfield  {journal} {\bibinfo  {journal} {Astron. Astrophys.}\ }\textbf {\bibinfo {volume} {339}},\ \bibinfo {pages} {165} (\bibinfo {year} {1998})}\BibitemShut {NoStop}%
\bibitem [{\citenamefont {Kebapc\i{}}\ \emph {et~al.}(2024)\citenamefont {Kebapc\i{}}, \citenamefont {Parlatan}, \citenamefont {Sert}, \citenamefont {\"Ozt\"urk}, \citenamefont {Ba\c{s}ar}, \citenamefont {\c{S}ahin}, \citenamefont {Bilir}, \citenamefont {Ferber}, \citenamefont {Tamanis},\ and\ \citenamefont {Kr\"oger}}]{Kebapci.2024}%
  \BibitemOpen
  \bibfield  {author} {\bibinfo {author} {\bibfnamefont {T.~Y.}\ \bibnamefont {Kebapc\i{}}}, \bibinfo {author} {\bibfnamefont {c.}~\bibnamefont {Parlatan}}, \bibinfo {author} {\bibfnamefont {S.}~\bibnamefont {Sert}}, \bibinfo {author} {\bibfnamefont {I.~K.}\ \bibnamefont {\"Ozt\"urk}}, \bibinfo {author} {\bibfnamefont {G.}~\bibnamefont {Ba\c{s}ar}}, \bibinfo {author} {\bibfnamefont {T.}~\bibnamefont {\c{S}ahin}}, \bibinfo {author} {\bibfnamefont {S.}~\bibnamefont {Bilir}}, \bibinfo {author} {\bibfnamefont {R.}~\bibnamefont {Ferber}}, \bibinfo {author} {\bibfnamefont {M.}~\bibnamefont {Tamanis}},\ and\ \bibinfo {author} {\bibfnamefont {S.}~\bibnamefont {Kr\"oger}},\ }\bibfield  {title} {\bibinfo {title} {Hyperfine {Structure} {Investigation} of {Singly} {Ionized} {Thulium} in {Fourier}-transform {Spectra}},\ }\href {https://doi.org/10.3847/1538-4357/ad47b7} {\bibfield  {journal} {\bibinfo  {journal} {Astrophys. J.}\ }\textbf {\bibinfo {volume} {970}},\ \bibinfo {pages} {23} (\bibinfo {year}
  {2024})}\BibitemShut {NoStop}%
\bibitem [{\citenamefont {Mansour}\ \emph {et~al.}(1989)\citenamefont {Mansour}, \citenamefont {Dinneen},\ and\ \citenamefont {Young}}]{Mansour.1989}%
  \BibitemOpen
  \bibfield  {author} {\bibinfo {author} {\bibfnamefont {N.}~\bibnamefont {Mansour}}, \bibinfo {author} {\bibfnamefont {T.}~\bibnamefont {Dinneen}},\ and\ \bibinfo {author} {\bibfnamefont {L.}~\bibnamefont {Young}},\ }\bibfield  {title} {\bibinfo {title} {High-precision measurements of hyperfine structure in {Tm} {II}, {N}$_2^+$ and {Sc} {II}},\ }\href {https://doi.org/10.1016/0168-583X(89)90972-5} {\bibfield  {journal} {\bibinfo  {journal} {Nucl. Instrum. Meth. Phys. Res. Sect. B}\ }\textbf {\bibinfo {volume} {40-41}},\ \bibinfo {pages} {252} (\bibinfo {year} {1989})}\BibitemShut {NoStop}%
\bibitem [{\citenamefont {Cheng}\ and\ \citenamefont {Childs}(1985)}]{Cheng.1985}%
  \BibitemOpen
  \bibfield  {author} {\bibinfo {author} {\bibfnamefont {K.~T.}\ \bibnamefont {Cheng}}\ and\ \bibinfo {author} {\bibfnamefont {W.~J.}\ \bibnamefont {Childs}},\ }\bibfield  {title} {\bibinfo {title} {Ab initio calculation of 4${f}^{N}$6${s}^{2}$ hyperfine structure in neutral rare-earth atoms},\ }\href {https://doi.org/10.1103/PhysRevA.31.2775} {\bibfield  {journal} {\bibinfo  {journal} {Phys. Rev. A}\ }\textbf {\bibinfo {volume} {31}},\ \bibinfo {pages} {2775} (\bibinfo {year} {1985})}\BibitemShut {NoStop}%
\bibitem [{\citenamefont {Bondarev}(2026)}]{Bondarev.2026}%
  \BibitemOpen
  \bibfield  {author} {\bibinfo {author} {\bibfnamefont {A.~I.}\ \bibnamefont {Bondarev}},\ }\bibfield  {title} {\bibinfo {title} {Calculation of {Hyperfine} {Structure} in {Tm} {II}},\ }\href {https://doi.org/10.3390/atoms14010007} {\bibfield  {journal} {\bibinfo  {journal} {Atoms}\ }\textbf {\bibinfo {volume} {14}},\ \bibinfo {pages} {7} (\bibinfo {year} {2026})}\BibitemShut {NoStop}%
\bibitem [{\citenamefont {Kramida}\ and\ \citenamefont {Ralchenko}(1999)}]{Kramida.1999}%
  \BibitemOpen
  \bibfield  {author} {\bibinfo {author} {\bibfnamefont {A.}~\bibnamefont {Kramida}}\ and\ \bibinfo {author} {\bibfnamefont {Y.}~\bibnamefont {Ralchenko}},\ }\href {https://doi.org/10.18434/T4W30F} {\bibinfo {title} {{NIST Atomic Spectra Database, NIST Standard Reference Database 78}}} (\bibinfo {year} {1999})\BibitemShut {NoStop}%
\bibitem [{\citenamefont {Kurucz}\ and\ \citenamefont {Bell}(1995)}]{Kurucz.1995}%
  \BibitemOpen
  \bibfield  {author} {\bibinfo {author} {\bibfnamefont {R.~L.}\ \bibnamefont {Kurucz}}\ and\ \bibinfo {author} {\bibfnamefont {B.}~\bibnamefont {Bell}},\ }\bibfield  {title} {\bibinfo {title} {Atomic line data},\ }\href {https://lweb.cfa.harvard.edu/amp/ampdata/kurucz23/sekur.html} {\bibfield  {journal} {\bibinfo  {journal} {Kurucz CD-ROM No. 23 Cambridge, Mass.: Smithsonian Astrophysical Observatory}\ } (\bibinfo {year} {1995})}\BibitemShut {NoStop}%
\bibitem [{\citenamefont {K{\"o}nig}\ \emph {et~al.}(2020{\natexlab{a}})\citenamefont {K{\"o}nig}, \citenamefont {Kr{\"a}mer}, \citenamefont {Geppert}, \citenamefont {Imgram}, \citenamefont {Maa{\ss}}, \citenamefont {Ratajczyk},\ and\ \citenamefont {N{\"o}rtersh{\"a}user}}]{Koenig.2020}%
  \BibitemOpen
  \bibfield  {author} {\bibinfo {author} {\bibfnamefont {K.}~\bibnamefont {K{\"o}nig}}, \bibinfo {author} {\bibfnamefont {J.}~\bibnamefont {Kr{\"a}mer}}, \bibinfo {author} {\bibfnamefont {C.}~\bibnamefont {Geppert}}, \bibinfo {author} {\bibfnamefont {P.}~\bibnamefont {Imgram}}, \bibinfo {author} {\bibfnamefont {B.}~\bibnamefont {Maa{\ss}}}, \bibinfo {author} {\bibfnamefont {T.}~\bibnamefont {Ratajczyk}},\ and\ \bibinfo {author} {\bibfnamefont {W.}~\bibnamefont {N{\"o}rtersh{\"a}user}},\ }\bibfield  {title} {\bibinfo {title} {{A new Collinear Apparatus for Laser Spectroscopy and Applied Science (COALA)}},\ }\href {https://doi.org/10.1063/5.0010903} {\bibfield  {journal} {\bibinfo  {journal} {Rev. Sci. Instrum.}\ }\textbf {\bibinfo {volume} {91}},\ \bibinfo {pages} {081301} (\bibinfo {year} {2020}{\natexlab{a}})}\BibitemShut {NoStop}%
\bibitem [{\citenamefont {Raeder}(2011)}]{Raeder.2011}%
  \BibitemOpen
  \bibfield  {author} {\bibinfo {author} {\bibfnamefont {S.}~\bibnamefont {Raeder}},\ }\emph {\bibinfo {title} {Spurenanalyse von {Aktiniden} in der {Umwelt} mittels {Resonanzionisations}-{Massenspektrometrie}}},\ \href {https://doi.org/10.25358/OPENSCIENCE-4788} {Ph.D. thesis},\ \bibinfo  {school} {Johannes Gutenberg-Universität Mainz} (\bibinfo {year} {2011})\BibitemShut {NoStop}%
\bibitem [{\citenamefont {Frömmgen}(2013)}]{Frommgen.2013}%
  \BibitemOpen
  \bibfield  {author} {\bibinfo {author} {\bibfnamefont {N.}~\bibnamefont {Frömmgen}},\ }\emph {\bibinfo {title} {Kollineare {Laserspektroskopie} an radioaktiven {Praseodymionen} und {Cadmiumatomen}}},\ \href {https://doi.org/10.25358/openscience-1784} {Ph.D. thesis},\ \bibinfo  {school} {Johannes Gutenberg-Universität Mainz} (\bibinfo {year} {2013})\BibitemShut {NoStop}%
\bibitem [{\citenamefont {Maaß}\ \emph {et~al.}(2020)\citenamefont {Maaß}, \citenamefont {König}, \citenamefont {Krämer}, \citenamefont {Miller}, \citenamefont {Minamisono}, \citenamefont {Nörtershäuser},\ and\ \citenamefont {Sommer}}]{Maass.2020}%
  \BibitemOpen
  \bibfield  {author} {\bibinfo {author} {\bibfnamefont {B.}~\bibnamefont {Maaß}}, \bibinfo {author} {\bibfnamefont {K.}~\bibnamefont {König}}, \bibinfo {author} {\bibfnamefont {J.}~\bibnamefont {Krämer}}, \bibinfo {author} {\bibfnamefont {A.~J.}\ \bibnamefont {Miller}}, \bibinfo {author} {\bibfnamefont {K.}~\bibnamefont {Minamisono}}, \bibinfo {author} {\bibfnamefont {W.}~\bibnamefont {Nörtershäuser}},\ and\ \bibinfo {author} {\bibfnamefont {F.}~\bibnamefont {Sommer}},\ }\href {https://doi.org/10.48550/arXiv.2007.02658} {\bibinfo {title} {A $4\pi$ fluorescence detection region for collinear laser spectroscopy}} (\bibinfo {year} {2020}),\ \Eprint {https://arxiv.org/abs/2007.02658} {arXiv:2007.02658 [physics.ins-det]} \BibitemShut {NoStop}%
\bibitem [{\citenamefont {M{\"u}ller}(2024)}]{Mueller.2024}%
  \BibitemOpen
  \bibfield  {author} {\bibinfo {author} {\bibfnamefont {P.~M.}\ \bibnamefont {M{\"u}ller}},\ }\emph {\bibinfo {title} {{Laserspectroscopic determination of the nuclear charge radius of $^{13\!}$C}}},\ \href {https://doi.org/10.26083/tuprints-00026746} {Ph.D. thesis},\ \bibinfo  {school} {Technische Universit{\"a}t Darmstadt}, \bibinfo {address} {Darmstadt} (\bibinfo {year} {2024})\BibitemShut {NoStop}%
\bibitem [{\citenamefont {Verlinde}\ \emph {et~al.}(2020)\citenamefont {Verlinde}, \citenamefont {Dockx}, \citenamefont {Geldhof}, \citenamefont {König}, \citenamefont {Studer}, \citenamefont {Cocolios}, \citenamefont {Groote}, \citenamefont {Ferrer}, \citenamefont {Kudryavtsev}, \citenamefont {Kieck}, \citenamefont {Moore}, \citenamefont {Nörtershäuser}, \citenamefont {Raeder}, \citenamefont {van~den Bergh}, \citenamefont {van Duppen},\ and\ \citenamefont {Wendt}}]{Verlinde.2020}%
  \BibitemOpen
  \bibfield  {author} {\bibinfo {author} {\bibfnamefont {M.}~\bibnamefont {Verlinde}}, \bibinfo {author} {\bibfnamefont {K.}~\bibnamefont {Dockx}}, \bibinfo {author} {\bibfnamefont {S.}~\bibnamefont {Geldhof}}, \bibinfo {author} {\bibfnamefont {K.}~\bibnamefont {König}}, \bibinfo {author} {\bibfnamefont {D.}~\bibnamefont {Studer}}, \bibinfo {author} {\bibfnamefont {T.~E.}\ \bibnamefont {Cocolios}}, \bibinfo {author} {\bibfnamefont {R.~P.}\ \bibnamefont {Groote}}, \bibinfo {author} {\bibfnamefont {R.}~\bibnamefont {Ferrer}}, \bibinfo {author} {\bibfnamefont {Y.}~\bibnamefont {Kudryavtsev}}, \bibinfo {author} {\bibfnamefont {T.}~\bibnamefont {Kieck}}, \bibinfo {author} {\bibfnamefont {I.}~\bibnamefont {Moore}}, \bibinfo {author} {\bibfnamefont {W.}~\bibnamefont {Nörtershäuser}}, \bibinfo {author} {\bibfnamefont {S.}~\bibnamefont {Raeder}}, \bibinfo {author} {\bibfnamefont {P.}~\bibnamefont {van~den Bergh}}, \bibinfo {author} {\bibfnamefont {P.}~\bibnamefont {van Duppen}},\ and\ \bibinfo {author}
  {\bibfnamefont {K.}~\bibnamefont {Wendt}},\ }\bibfield  {title} {\bibinfo {title} {On the performance of wavelength meters: {Part} 1—consequences for medium-to-high-resolution laser spectroscopy},\ }\href {https://doi.org/10.1007/s00340-020-07425-4} {\bibfield  {journal} {\bibinfo  {journal} {Appl. Phys. B}\ }\textbf {\bibinfo {volume} {126}},\ \bibinfo {pages} {85} (\bibinfo {year} {2020})}\BibitemShut {NoStop}%
\bibitem [{\citenamefont {K{\"o}nig}\ \emph {et~al.}(2020{\natexlab{b}})\citenamefont {K{\"o}nig}, \citenamefont {Imgram}, \citenamefont {Kr{\"a}mer}, \citenamefont {Maa{\ss}}, \citenamefont {Mohr}, \citenamefont {Ratajczyk}, \citenamefont {Sommer},\ and\ \citenamefont {N{\"o}rtersh{\"a}user}}]{Koenig.2020b}%
  \BibitemOpen
  \bibfield  {author} {\bibinfo {author} {\bibfnamefont {K.}~\bibnamefont {K{\"o}nig}}, \bibinfo {author} {\bibfnamefont {P.}~\bibnamefont {Imgram}}, \bibinfo {author} {\bibfnamefont {J.}~\bibnamefont {Kr{\"a}mer}}, \bibinfo {author} {\bibfnamefont {B.}~\bibnamefont {Maa{\ss}}}, \bibinfo {author} {\bibfnamefont {K.}~\bibnamefont {Mohr}}, \bibinfo {author} {\bibfnamefont {T.}~\bibnamefont {Ratajczyk}}, \bibinfo {author} {\bibfnamefont {F.}~\bibnamefont {Sommer}},\ and\ \bibinfo {author} {\bibfnamefont {W.}~\bibnamefont {N{\"o}rtersh{\"a}user}},\ }\bibfield  {title} {\bibinfo {title} {On the performance of wavelength meters: Part 2 -- frequency-comb based characterization for more accurate absolute wavelength determinations},\ }\href {https://doi.org/10.1007/s00340-020-07433-4} {\bibfield  {journal} {\bibinfo  {journal} {Appl. Phys. B}\ }\textbf {\bibinfo {volume} {126}},\ \bibinfo {pages} {86} (\bibinfo {year} {2020}{\natexlab{b}})}\BibitemShut {NoStop}%
\bibitem [{\citenamefont {König}\ \emph {et~al.}(2024)\citenamefont {König}, \citenamefont {Köhler}, \citenamefont {Palmes}, \citenamefont {Badura}, \citenamefont {Dockery}, \citenamefont {Minamisono}, \citenamefont {Meisner}, \citenamefont {Müller}, \citenamefont {Nörtershäuser},\ and\ \citenamefont {Passon}}]{Koenig.2024}%
  \BibitemOpen
  \bibfield  {author} {\bibinfo {author} {\bibfnamefont {K.}~\bibnamefont {König}}, \bibinfo {author} {\bibfnamefont {F.}~\bibnamefont {Köhler}}, \bibinfo {author} {\bibfnamefont {J.}~\bibnamefont {Palmes}}, \bibinfo {author} {\bibfnamefont {H.}~\bibnamefont {Badura}}, \bibinfo {author} {\bibfnamefont {A.}~\bibnamefont {Dockery}}, \bibinfo {author} {\bibfnamefont {K.}~\bibnamefont {Minamisono}}, \bibinfo {author} {\bibfnamefont {J.}~\bibnamefont {Meisner}}, \bibinfo {author} {\bibfnamefont {P.}~\bibnamefont {Müller}}, \bibinfo {author} {\bibfnamefont {W.}~\bibnamefont {Nörtershäuser}},\ and\ \bibinfo {author} {\bibfnamefont {S.}~\bibnamefont {Passon}},\ }\bibfield  {title} {\bibinfo {title} {High voltage determination and stabilization for collinear laser spectroscopy applications},\ }\href {https://doi.org/10.1063/5.0218649} {\bibfield  {journal} {\bibinfo  {journal} {Rev. Sci. Instrum.}\ }\textbf {\bibinfo {volume} {95}},\ \bibinfo {pages} {083307} (\bibinfo {year} {2024})}\BibitemShut {NoStop}%
\bibitem [{\citenamefont {König}\ \emph {et~al.}(2021)\citenamefont {König}, \citenamefont {Minamisono}, \citenamefont {Lantis}, \citenamefont {Pineda},\ and\ \citenamefont {Powel}}]{Konig.2021}%
  \BibitemOpen
  \bibfield  {author} {\bibinfo {author} {\bibfnamefont {K.}~\bibnamefont {König}}, \bibinfo {author} {\bibfnamefont {K.}~\bibnamefont {Minamisono}}, \bibinfo {author} {\bibfnamefont {J.}~\bibnamefont {Lantis}}, \bibinfo {author} {\bibfnamefont {S.}~\bibnamefont {Pineda}},\ and\ \bibinfo {author} {\bibfnamefont {R.}~\bibnamefont {Powel}},\ }\bibfield  {title} {\bibinfo {title} {Beam energy determination via collinear laser spectroscopy},\ }\href {https://doi.org/10.1103/PhysRevA.103.032806} {\bibfield  {journal} {\bibinfo  {journal} {Phys. Rev. A}\ }\textbf {\bibinfo {volume} {103}},\ \bibinfo {pages} {032806} (\bibinfo {year} {2021})}\BibitemShut {NoStop}%
\bibitem [{\citenamefont {Imgram}\ \emph {et~al.}(2019)\citenamefont {Imgram}, \citenamefont {K{\"o}nig}, \citenamefont {Kr{\"a}mer}, \citenamefont {Ratajczyk}, \citenamefont {M{\"u}ller}, \citenamefont {Surzhykov},\ and\ \citenamefont {N{\"o}rtersh{\"a}user}}]{Imgram.2019}%
  \BibitemOpen
  \bibfield  {author} {\bibinfo {author} {\bibfnamefont {P.}~\bibnamefont {Imgram}}, \bibinfo {author} {\bibfnamefont {K.}~\bibnamefont {K{\"o}nig}}, \bibinfo {author} {\bibfnamefont {J.}~\bibnamefont {Kr{\"a}mer}}, \bibinfo {author} {\bibfnamefont {T.}~\bibnamefont {Ratajczyk}}, \bibinfo {author} {\bibfnamefont {R.~A.}\ \bibnamefont {M{\"u}ller}}, \bibinfo {author} {\bibfnamefont {A.}~\bibnamefont {Surzhykov}},\ and\ \bibinfo {author} {\bibfnamefont {W.}~\bibnamefont {N{\"o}rtersh{\"a}user}},\ }\bibfield  {title} {\bibinfo {title} {{Collinear laser spectroscopy at ion-trap accuracy: Transition frequencies and isotope shifts in the $6s\,^{2}\mathrm{S}_{1/2} \rightarrow 6p\,^2\mathrm{P}_{1/2,3/2}$ transitions in Ba$^+$}},\ }\href {https://doi.org/10.1103/PhysRevA.99.012511} {\bibfield  {journal} {\bibinfo  {journal} {Phys. Rev. A}\ }\textbf {\bibinfo {volume} {99}},\ \bibinfo {pages} {012511} (\bibinfo {year} {2019})}\BibitemShut {NoStop}%
\bibitem [{\citenamefont {M{\"u}ller}\ \emph {et~al.}(2020)\citenamefont {M{\"u}ller}, \citenamefont {K{\"o}nig}, \citenamefont {Imgram}, \citenamefont {Kr{\"a}mer},\ and\ \citenamefont {N{\"o}rtersh{\"a}user}}]{Muller.2020}%
  \BibitemOpen
  \bibfield  {author} {\bibinfo {author} {\bibfnamefont {P.}~\bibnamefont {M{\"u}ller}}, \bibinfo {author} {\bibfnamefont {K.}~\bibnamefont {K{\"o}nig}}, \bibinfo {author} {\bibfnamefont {P.}~\bibnamefont {Imgram}}, \bibinfo {author} {\bibfnamefont {J.}~\bibnamefont {Kr{\"a}mer}},\ and\ \bibinfo {author} {\bibfnamefont {W.}~\bibnamefont {N{\"o}rtersh{\"a}user}},\ }\bibfield  {title} {\bibinfo {title} {{Collinear laser spectroscopy of Ca$^+$: Solving the field-shift puzzle of the $4s \,^2\mathrm{S}_{1/2} \rightarrow 4p\,^2\mathrm{P}_{1/2,3/2}$ transitions}},\ }\href {https://doi.org/10.1103/PhysRevResearch.2.043351} {\bibfield  {journal} {\bibinfo  {journal} {Phys. Rev. Res.}\ }\textbf {\bibinfo {volume} {2}},\ \bibinfo {pages} {043351} (\bibinfo {year} {2020})}\BibitemShut {NoStop}%
\bibitem [{\citenamefont {Palmes}\ \emph {et~al.}(2026)\citenamefont {Palmes}, \citenamefont {König}, \citenamefont {Sahoo}, \citenamefont {Bodnar}, \citenamefont {Candiello}, \citenamefont {Dorne}, \citenamefont {Groote}, \citenamefont {Imgram}, \citenamefont {Lopp}, \citenamefont {Müller}, \citenamefont {Nörtershäuser}, \citenamefont {Ohayon},\ and\ \citenamefont {Duyse}}]{Palmes.2026}%
  \BibitemOpen
  \bibfield  {author} {\bibinfo {author} {\bibfnamefont {J.}~\bibnamefont {Palmes}}, \bibinfo {author} {\bibfnamefont {K.}~\bibnamefont {König}}, \bibinfo {author} {\bibfnamefont {B.~K.}\ \bibnamefont {Sahoo}}, \bibinfo {author} {\bibfnamefont {H.}~\bibnamefont {Bodnar}}, \bibinfo {author} {\bibfnamefont {A.}~\bibnamefont {Candiello}}, \bibinfo {author} {\bibfnamefont {A.}~\bibnamefont {Dorne}}, \bibinfo {author} {\bibfnamefont {R.~d.}\ \bibnamefont {Groote}}, \bibinfo {author} {\bibfnamefont {P.}~\bibnamefont {Imgram}}, \bibinfo {author} {\bibfnamefont {I.}~\bibnamefont {Lopp}}, \bibinfo {author} {\bibfnamefont {P.}~\bibnamefont {Müller}}, \bibinfo {author} {\bibfnamefont {W.}~\bibnamefont {Nörtershäuser}}, \bibinfo {author} {\bibfnamefont {B.}~\bibnamefont {Ohayon}},\ and\ \bibinfo {author} {\bibfnamefont {R.~V.}\ \bibnamefont {Duyse}},\ }\bibfield  {title} {\bibinfo {title} {Nuclear charge radii of {Sr} isotopes: Reevaluation based on transition frequency measurements in the $5s$-$5p$-$4d$ manifold in
  {Sr}${}^{+}$},\ }\bibfield  {journal} {\bibinfo  {journal} {Phys. Rev. Res.}\ }\href {https://doi.org/10.1103/th38-y1t9} {10.1103/th38-y1t9} (\bibinfo {year} {2026})\BibitemShut {NoStop}%
\bibitem [{\citenamefont {Imgram}\ \emph {et~al.}(2023)\citenamefont {Imgram}, \citenamefont {K\"onig}, \citenamefont {Maa\ss{}}, \citenamefont {M\"uller},\ and\ \citenamefont {N\"ortersh\"auser}}]{Imgram.2023}%
  \BibitemOpen
  \bibfield  {author} {\bibinfo {author} {\bibfnamefont {P.}~\bibnamefont {Imgram}}, \bibinfo {author} {\bibfnamefont {K.}~\bibnamefont {K\"onig}}, \bibinfo {author} {\bibfnamefont {B.}~\bibnamefont {Maa\ss{}}}, \bibinfo {author} {\bibfnamefont {P.}~\bibnamefont {M\"uller}},\ and\ \bibinfo {author} {\bibfnamefont {W.}~\bibnamefont {N\"ortersh\"auser}},\ }\bibfield  {title} {\bibinfo {title} {{Collinear Laser Spectroscopy of $2\text{ }^{3}{S}_{1}\ensuremath{\rightarrow}2\text{ }^{3}{P}_{J}$ Transitions in Helium-like ${^{12}\mathrm{C}}^{4+}$}},\ }\href {https://doi.org/10.1103/PhysRevLett.131.243001} {\bibfield  {journal} {\bibinfo  {journal} {Phys. Rev. Lett.}\ }\textbf {\bibinfo {volume} {131}},\ \bibinfo {pages} {243001} (\bibinfo {year} {2023})}\BibitemShut {NoStop}%
\bibitem [{\citenamefont {König}\ \emph {et~al.}(2026)\citenamefont {König}, \citenamefont {Müller}, \citenamefont {Gesser}, \citenamefont {Burbach}, \citenamefont {Gandolfi}, \citenamefont {Heinz}, \citenamefont {Imgram}, \citenamefont {Lovato}, \citenamefont {Maris}, \citenamefont {Miyagi}, \citenamefont {Nörtershäuser}, \citenamefont {Roth}, \citenamefont {Spahn},\ and\ \citenamefont {Schwenk}}]{Koenig.2026}%
  \BibitemOpen
  \bibfield  {author} {\bibinfo {author} {\bibfnamefont {K.}~\bibnamefont {König}}, \bibinfo {author} {\bibfnamefont {P.}~\bibnamefont {Müller}}, \bibinfo {author} {\bibfnamefont {T.}~\bibnamefont {Gesser}}, \bibinfo {author} {\bibfnamefont {E.}~\bibnamefont {Burbach}}, \bibinfo {author} {\bibfnamefont {S.}~\bibnamefont {Gandolfi}}, \bibinfo {author} {\bibfnamefont {M.}~\bibnamefont {Heinz}}, \bibinfo {author} {\bibfnamefont {P.}~\bibnamefont {Imgram}}, \bibinfo {author} {\bibfnamefont {A.}~\bibnamefont {Lovato}}, \bibinfo {author} {\bibfnamefont {P.}~\bibnamefont {Maris}}, \bibinfo {author} {\bibfnamefont {T.}~\bibnamefont {Miyagi}}, \bibinfo {author} {\bibfnamefont {W.}~\bibnamefont {Nörtershäuser}}, \bibinfo {author} {\bibfnamefont {R.}~\bibnamefont {Roth}}, \bibinfo {author} {\bibfnamefont {J.}~\bibnamefont {Spahn}},\ and\ \bibinfo {author} {\bibfnamefont {A.}~\bibnamefont {Schwenk}},\ }\bibfield  {title} {\bibinfo {title} {{Accurate charge radius measurement of ${}^{14}$C confronts \textit{ab initio}
  theory}},\ }\bibfield  {journal} {\bibinfo  {journal} {Phys. Rev. Lett.}\ }\href {https://doi.org/10.1103/f5s7-1t7w} {10.1103/f5s7-1t7w} (\bibinfo {year} {2026})\BibitemShut {NoStop}%
\bibitem [{\citenamefont {Meggers}\ \emph {et~al.}(1975)\citenamefont {Meggers}, \citenamefont {Corliss},\ and\ \citenamefont {Scribner}}]{Meggers.1975}%
  \BibitemOpen
  \bibfield  {author} {\bibinfo {author} {\bibfnamefont {W.~F.}\ \bibnamefont {Meggers}}, \bibinfo {author} {\bibfnamefont {C.~H.}\ \bibnamefont {Corliss}},\ and\ \bibinfo {author} {\bibfnamefont {B.~F.}\ \bibnamefont {Scribner}},\ }\href {https://doi.org/10.6028/NBS.MONO.145p1} {\emph {\bibinfo {title} {{Tables of Spectral-Line Intensities, Part I – Arranged by Elements, Part II – Arranged by Wavelengths}}}}\ (\bibinfo  {publisher} {Nat. Bur. Stand. Monograph 145},\ \bibinfo {year} {1975})\BibitemShut {NoStop}%
\bibitem [{\citenamefont {Reader}\ \emph {et~al.}(1980)\citenamefont {Reader}, \citenamefont {Corliss}, \citenamefont {Wiese},\ and\ \citenamefont {Martin}}]{Reader.1980}%
  \BibitemOpen
  \bibfield  {author} {\bibinfo {author} {\bibfnamefont {J.}~\bibnamefont {Reader}}, \bibinfo {author} {\bibfnamefont {C.~H.}\ \bibnamefont {Corliss}}, \bibinfo {author} {\bibfnamefont {W.~L.}\ \bibnamefont {Wiese}},\ and\ \bibinfo {author} {\bibfnamefont {G.~A.}\ \bibnamefont {Martin}},\ }\href {https://doi.org/10.6028/NBS.NSRDS.68} {\emph {\bibinfo {title} {{Wavelengths and Transition Probabilities for Atoms and Atomic Ions Part I. Wavelengths}}}}\ (\bibinfo  {publisher} {Nat. Bur. Stand., NSRDS-NBS 68},\ \bibinfo {year} {1980})\BibitemShut {NoStop}%
\bibitem [{\citenamefont {Bodnar}\ \emph {et~al.}(2026)\citenamefont {Bodnar}, \citenamefont {König},\ and\ \citenamefont {Nörtershäuser}}]{data}%
  \BibitemOpen
  \bibfield  {author} {\bibinfo {author} {\bibfnamefont {H.}~\bibnamefont {Bodnar}}, \bibinfo {author} {\bibfnamefont {K.}~\bibnamefont {König}},\ and\ \bibinfo {author} {\bibfnamefont {W.}~\bibnamefont {Nörtershäuser}},\ }\bibfield  {title} {\bibinfo {title} {{Data: Measurement Data and Analysis of Collinear Laser Spectroscopy of stable Tm II at COALA}},\ }\href {https://doi.org/10.48328/tudatalib-2347} {10.48328/tudatalib-2347} (\bibinfo {year} {2026})\BibitemShut {NoStop}%
\end{thebibliography}
\end{document}